\documentclass[11pt,a4paper]{article}

\usepackage[english]{babel}

\usepackage[most]{tcolorbox}

\usepackage{algorithm}
\usepackage{algpseudocode}

\usepackage[utf8]{inputenc}
\usepackage[T1]{fontenc}
\usepackage{microtype}
\usepackage{moresize}

\usepackage{csquotes,textcase,xspace}

\usepackage{geometry}
\usepackage{setspace}
\usepackage{graphicx}
\usepackage{adjustbox} 
\usepackage{xcolor}
\usepackage[font=small,labelfont=bf,labelsep=space]{caption}
\usepackage{subfigure}
\addto\captionsenglish{}
\addto\captionsenglish{}
\definecolor{JoliBleu}{rgb}{0,0.55,0.55}
\definecolor{JoliVert}{rgb}{0.15,0.6,0}
\definecolor{JoliRouge}{rgb}{0.86,0.08,0}
\definecolor{JoliJaune}{rgb}{1,0.75,0}
\definecolor{JoliGris}{rgb}{0.52,0.52,0.51}
\definecolor{myblue}{RGB}{26, 77, 116}
\definecolor{myorange}{RGB}{181, 116, 30}
\definecolor{mydarkorange}{RGB}{166, 88, 0}
\definecolor{mygreen}{RGB}{21, 124, 80}
\definecolor{myblack}{RGB}{43, 65, 82}
\definecolor{myred}{rgb}{0.5, 0.0, 0.13}
\definecolor{clusteringred1}{RGB}{139, 0, 0}
\definecolor{clusteringred2}{RGB}{231, 76, 60}
\definecolor{clusteringyellow1}{RGB}{192, 200, 44}

\usepackage{titlesec}
\titleformat{\section}[block]{\Large\boldmath\bfseries}{\thesection}{1em}{}
\titleformat{\subsection}[block]{\large\boldmath\bfseries}{\thesubsection}{0.5em}{}
\usepackage{appendix}

\makeatletter

\@addtoreset{equation}{section}
\makeatother

\usepackage[bottom]{footmisc}
\usepackage{fancyhdr}

\usepackage{titletoc}
\usepackage[nosort]{cite}
\usepackage{hyperref}
\hypersetup{colorlinks=true,
        pdfstartview=FitV,
        linkcolor= mydarkorange,
        citecolor= mydarkorange, 
        urlcolor= JoliGris!60!black,
        hypertexnames=false,
        linktoc=page}

\usepackage{tikz}
\usepackage{tcolorbox}
\usetikzlibrary{calc}
\usetikzlibrary{shapes.geometric, arrows.meta, positioning}

\definecolor{deepblue}{RGB}{41, 128, 185}
\definecolor{lightblue}{RGB}{174, 214, 241}
\definecolor{emerald}{RGB}{46, 204, 113}
\definecolor{lightemerald}{RGB}{212, 245, 227}
\definecolor{coral}{RGB}{231, 76, 60}
\definecolor{lightcoral}{RGB}{250, 215, 212}
\definecolor{charcoal}{RGB}{52, 73, 94}
\definecolor{lightgray}{RGB}{150, 150, 150}

\tikzstyle{startstop} = [
ellipse, 
minimum width=3cm, 
minimum height=1cm, 
text centered, 
draw=charcoal, 
fill=lightgray,
line width=1.2pt,
font=\sffamily\small\bfseries
]

\tikzstyle{process} = [
rectangle, 
minimum width=5.5cm, 
minimum height=1.1cm, 
text centered,
draw=myblue, 
fill=lightblue,
line width=1pt,
rounded corners=2pt,
font=\sffamily\small
]

\tikzstyle{decision} = [
rectangle, 
minimum width=3.2cm, 
minimum height=1.cm, 
text centered, 
draw=myorange, 
fill=myorange!20,
line width=1pt,
font=\sffamily\small\bfseries,
aspect=1.3
]

\tikzstyle{arrow} = [
thick,
->,
>=Stealth,
color=charcoal,
line width=1.5pt
]

\tikzstyle{arrow_label} = [
font=\sffamily\small\bfseries,
color=charcoal
]

\usepackage{amsmath,amssymb,amsfonts,dsfont}
\usepackage{mathrsfs}
\usepackage{physics}
\usepackage{ytableau}
\ytableausetup{boxsize=1.1em,centertableaux}
\usepackage{stmaryrd}
\usepackage{nicefrac}
\allowdisplaybreaks[1]
\usepackage{cases}
\usepackage{bm}
\usepackage{bbm}

\usepackage{multirow}
\usepackage{booktabs}
\usepackage{pdflscape}
\usepackage{array}
\usepackage{arydshln}

\usepackage[shortlabels]{enumitem}

\renewcommand{\d}{\ensuremath{\mathrm{d}}\xspace}

\def\sst#1{{\scriptscriptstyle #1}}

\newcommand{\bK}{{\bar{K}}}
\newcommand{\bL}{{\bar{L}}}
\newcommand{\bM}{{\bar{M}}}
\newcommand{\bN}{{\bar{N}}}
\newcommand{\bP}{{\bar{P}}}
\newcommand{\bQ}{{\bar{Q}}}
\newcommand{\bR}{{\bar{R}}}
\newcommand{\bS}{{\bar{S}}}

\newcommand{\bzero}{{\bar{0}}}
\newcommand{\balpha}{{\bar{\alpha}}}
\newcommand{\bbeta}{{\bar{\beta}}}

\def\sst#1{{\scriptscriptstyle #1}}

\def\0{{\sst{(0)}}}
\def\1{{\sst{(1)}}}
\def\2{{\sst{(2)}}}
\def\3{{\sst{(3)}}}
\def\4{{\sst{(4)}}}
\def\5{{\sst{(5)}}}
\def\6{{\sst{(6)}}}
\def\7{{\sst{(7)}}}
\def\8{{\sst{(8)}}}

\newcommand{\be}{\begin{equation}}
\newcommand{\ee}{\end{equation}}

\begin{document}

\begin{titlepage}

\begin{flushright}

\today
\end{flushright}

\vspace{25pt}

   \begin{center}
   \baselineskip=16pt

{\Large Machine Learning the Conformal Manifold of Holographic CFT$_{2}$’s}

\vspace{25pt}

{\large  Bastien Duboeuf$^{1}$, Camille Eloy$^{2}$ \,and\, Gabriel Larios$^{3}$}
		
\vspace{25pt}

	\begin{small}

	{\it $^{1}$ Max-Planck-Institut f\"ur Gravitationsphysik (Albert-Einstein-Institut)\\ Am M\"uhlenberg 1, DE-14476 Potsdam, Germany}  \\

	\vspace{10pt}

	{\it $^{2}$ ENS de Lyon, CNRS, LPENSL, UMR5672,\\ 69342, Lyon cedex 07, France}  \\

	\vspace{10pt}
	
	{\it $^{3}$ Mitchell Institute for Fundamental Physics and Astronomy, \\
	Texas A\&M University, College Station, TX, 77843, USA}     \\
		
	\end{small}

\vskip 50pt

\end{center}

\begin{abstract}
	We investigate the structure of conformal manifolds around AdS$_3 \times S^3$ which lift from continuous flat directions in the scalar potential of gauged supergravity resulting from six-dimensional $\mathcal{N}=(1,1)$ supergravity.
	Our approach combines numerical exploration and symbolic inference. For the latter, we develop a symbolic regression algorithm based on Annealed Sequential Monte Carlo samplers, a combination of Annealed Importance Sampling and Sequential Monte Carlo samplers, well-suited to uncovering polynomial constraints in high-dimensional parameter spaces. 
	The algorithm reconstructs a set of polynomial relations that provides an explicit analytic parametrization of a new family of solutions. 
\end{abstract}

\vfill

\end{titlepage}

\tableofcontents

\section{Introduction}

The AdS/CFT correspondence stands as one of the most profound dualities in theoretical physics, establishing a remarkable equivalence between string theory solutions on anti-de Sitter (AdS) spacetimes and conformal field theories (CFTs) living on their boundaries~\cite{Maldacena:1997re}. 
This correspondence has revolutionized our understanding of both quantum gravity and strongly coupled field theories, providing unprecedented insights into the holographic nature of gravity. 
Within this holographic framework, supergravity theories in AdS backgrounds serve as the low-energy effective descriptions of string theory compactifications, making them natural laboratories for exploring the gravitational side of the duality. On the other side, CFT’s appear in multiple scenarios, providing nice descriptions of phase transitions in statistical physics, and from a fundamental perspective, they are of utmost importance in quantum field theories (QFTs), where they describe fixed points of Renormalisation Group flow. 
From the latter perspective, an interesting question is  whether CFTs are isolated fixed points or instead belong to a continuous families. 
In that case, the space of continuous deformations that takes from one CFT to another is called a conformal manifold. 
These deformations are parametrised by exactly marginal operators, or in other words, operators that preserve the conformal symmetry, \textit{i.e.} whose $\beta$ functions exactly vanish.

From the AdS/CFT correspondence perspective, conformal manifolds on the boundary theory are dual to continuous families of AdS solutions where the (possibly warped) AdS factor stays undeformed. If a consistent truncation on these solutions exists, conformal manifolds can also be identified as flat directions in the scalar potential of the truncation. Along these directions, the scalar field configurations vary continuously while the cosmological constant remains unchanged. Supersymmetry is believed to be necessary for the existence of holographic conformal manifolds, as non-supersymmetric AdS solutions are expected to be unstable~\cite{Ooguri:2016pdq,Palti:2019pca}. However, recent investigations have identified AdS$_4$ configurations that appear to evade this requirement, with no evidence of standard decay channels -- neither perturbative nor non-perturbative -- being present~\cite{Giambrone:2021wsm}. The situation is even richer in the context of AdS$_3$/CFT$_2$, as current-current deformations, given by products $J\bar J$ of (anti-)holomorphic conserved currents, are exactly marginal in two dimensions~\cite{Chaudhuri:1988qb}, although they may break supersymmetry. The question of the exact form of the gravity dual to $J\bar J$ deformations is however open. A well-known example is given in ref.~\cite{Aharony:2001dp,Dong:2014tsa}, and recent works have demonstrated the existence of vast families of classically marginal deformations of the ${\rm AdS}_{3}\times S^{3}\times {\rm T}^{4}$ and ${\rm AdS}_{3}\times S^{3}\times S^{3}\times S^{1}$ solutions of type IIB supergravity that are perturbatively stable despite supersymmetry breaking~\cite{Eloy:2023zzh,Eloy:2023acy,Eloy:2024lwn}. These deformations were shown in ref.~\cite{Eloy:2024lwn} to be equivalent to current-current deformations of the worldsheet Wess-Zumino-Witten models~\cite{Gepner:1986wi} describing the undeformed solutions~\cite{Eberhardt:2017pty,Eberhardt:2018ouy}, giving strong indications that their CFT duals arise from $J\bar J$ deformations as well.

Although extensive, the families of deformations found in the aforementioned papers have not exhausted all possibilities for marginal deformations of three-dimensional supergravity solutions. A complete classification would require a systematic study of the flat directions in the supergravity scalar potential $V$ that defines the AdS configurations, as those directions correspond to classical marginal deformations of the holographic CFT in the large $N$ limit. These flat directions are continuous sets in the space of scalar fields along which the value of the potential stay fixed, \textit{i.e.}  along which $\nabla V = 0$. However, the explicit characterisation of flat directions presents formidable technical challenges. Supergravity scalar potentials, even in truncated models, typically involve dozens of scalar fields with intricate non-linear interactions. The resulting expressions for critical points -- where all first-order derivatives vanish -- quickly become too complex for traditional symbolic manipulation, rendering analytical approaches computationally intractable.

The emergence of machine learning techniques opens new avenues for addressing such complex problems in theoretical physics. Instead of solving the full symbolic system analytically from the outset, one can employ numerical methods to sample the solution space and subsequently apply symbolic regression techniques to extract analytical patterns from the data. This hybrid approach has the potential to bypass the computational bottlenecks inherent to purely symbolic methods, while still having the potential to uncover exact analytical expressions.

Machine learning strategies have previously been applied to identify new isolated vacua in $\mathrm{SO}(8)$ supergravity \cite{Comsa:2019rcz,Berman:2022jqn}. More broadly, there has been increasing interest in applying machine learning and numerical techniques across various domains of high-energy physics. This includes, for instance, the characterisation of Calabi–Yau metrics and hypersurfaces \cite{Ashmore:2019wzb,Berman:2021mcw,Larfors:2022nep,Berglund:2022gvm,Jejjala:2020wcc,Douglas:2006rr,Larfors:2021pbb,He:2018jtw}, as well as broader efforts to explore the string theory landscape \cite{He:2017aed,Carifio:2017bov,Ruehle:2017mzq,Walden:2025cpf}. Additional applications include studies in CFT \cite{Chen:2020dxg}, investigations of the supergravity landscape \cite{Brady:2025zzi,Krishnan:2020sfg}, and explorations of the AdS/CFT correspondence \cite{Hashimoto:2018ftp}. More generally, machine learning has found utility in the study of string theory, geometry, and fundamental physics \cite{Ruehle:2020jrk,He:2023csq,Bao:2021auj}.

In this work, we demonstrate the viability of the aforementioned machine learning approach by applying it to a five-scalar subsector of a 13 scalar consistent truncation of six-dimensional non-chiral $\mathcal{N} = (1,1)$ supergravity on ${\rm AdS}_3 \times S^3$, or to type IIB supergravity on ${\rm AdS}_3 \times S^3 \times T^4$. Our methodology combines gradient descent sampling of the conformal manifold with a symbolic regression technique. There exists a large body of literature on symbolic regression, using methods from genetic programming \cite{koza1994genetic,virgolin2021improving,randall2022bingo,burlacu2019parsimony}, to deep learning \cite{petersen2019deep,kamienny2022end}, generative models \cite{valipour2021symbolicgpt}, diffusion models \cite{bastiani2025diffusion}, and equation learning with the nodes being symbolic operations \cite{2018arXiv180607259S}. A state-of-the-art algorithm is AIFeynman \cite{Udrescu:2019mnk}. It uses neural networks to identify structures in the dataset (such as translational symmetries, multiplicative separability, compositionality\dots) to recursively define simpler problems on which they can fit the solutions with polynomials and basic functions.  However, this method is slow, usually limited to low-dimensional spaces, and, by construction, can only fit one expression at a time. Alternatively, we develop here a symbolic regression algorithm based on an Annealed Sequential Monte Carlo sampler (ASMC)~\cite{zimmermann2021nested,syed2024optimisedannealedsequentialmonte} (combining the Annealed Importance Sampling~\cite{neal1998annealedimportancesampling} approach with a Sequential Monte Carlo sampler (SMC)~\cite{del2006sequential}), since this approach is much better suited to the study of geometric loci defined by intersecting polynomials in high-dimensional spaces.

The study of the five-parameter potential is separated into several parts. We first use a gradient descent to efficiently sample the underlying conformal manifold. Combined with numerical analyses, including a local principal component analysis, we can identify the existence of a three-dimensional continuous family. We then use an ASMC technique in order to do symbolic regression. As we will indeed demonstrate in the paper, there exist polynomial constraints on the 5 parameters, viewed as embedding coordinates, that project them onto the 3 dimensional manifolds. We manage to identify 8 of those constraints on the data, not all independent, which once solved  provide an explicit three-dimensional family of solutions.

\paragraph{}
The paper is organized as follows. In sec.~\ref{sec:sugra}, we establish the supergravity setup, presenting the scalar potential in its full complexity and motivating the restriction to five fields. Sec.~\ref{Sec:AIS-SMC} details our Annealed Sequential Monte Carlo sampler approach to symbolic regression. Our main results, both numerical an analytical, are presented in sec.~\ref{sec:results}, and sec.~\ref{sec:sugrasol} gives some details on the new supergravity solutions discovered by the numerical analysis. We conclude with prospects for extending this approach to higher-dimensional cases and its broader implications for systematic studies of conformal manifolds in holographic theories.

\section{Supergravity Setup} \label{sec:sugra}
Three-dimensional $\mathcal{N}=8$ (half-maximal) gauged supergravity is governed by the Lagrangian~\cite{Nicolai:2001ac,deWit:2003ja}
\begin{equation}	\label{eq: lagrangian_rephrased} 
	e^{-1}\mathcal{L}=R+\frac1{8}g^{\mu\nu}D_\mu M^{\bM\bN}D_\nu M_{\bM\bN}+e^{-1}\mathcal{L}_{\text{CS}}-V,
\end{equation}
which comprises an Einstein-Hilbert term $R$, a kinetic term for scalar fields parametrised by the matrix $M_{\bM\bN}$, a Chern-Simons contribution $\mathcal{L}_{\text{CS}}$, and a scalar potential $V$, see app.~\ref{app:sugra} for details on the definition of the terms. The scalar degrees of freedom parametrise the coset space
\begin{equation}	\label{eq: scalarcoset_rephrased}
	\frac{\text{SO}(8,4)}{\text{SO}(8)\times\text{SO}(4)},
\end{equation}
through the symmetric matrix $M_{\bK\bL}$.\footnote{In full generality, the coset space of half-maximal supergravity in three dimensions is $\text{SO}(8,p)/\big(\text{SO}(8)\times\text{SO}(p)\big)$. We consider here the case $p=4$ only.} With appropriate gauging (see eq.~\eqref{eq:embedtens}), this $\text{SO}(8,4)$ theory is a consistent truncation of six-dimensional ${\cal N}=(1,1)$ supergravity on $S^{3}$~\cite{Cvetic:2000dm,Deger:2014ofa,Hohm:2017wtr}. It features an ${\rm AdS}_{3}$ stationary point at the scalar origin ($M_{\bK\bL}=\delta_{\bK\bL}$, the ${\rm SO}(8,4)$ identity matrix), corresponding to an ${\rm AdS}_{3}\times S^{3}$ solution in 6d. In the following, we will be interested in exploring flat directions of the potential $V$ around this point, \textit{i.e.} stationary points that are continuously connected to the origin. In three dimensions, the flat directions constitute a family of ${\rm AdS}_{3}$ vacua sharing the same cosmological constant. The corresponding solutions in six dimensions are of the form ${\rm AdS}_{3}\times M^{3}$, with $M^{3}$ some deformation of the round sphere $S^{3}$, parametrised by the constant scalar vevs defining the 3d vacua.

We parametrise the scalars of the theory following ref.~\cite{Eloy:2021fhc} (see also app.~\ref{app:sugra} for some details): 13 scalars parametrised by a symmetric ${\rm GL}(3,\mathbb{R})$ matrix $m = \nu\nu^{T}$ where
\begin{equation} \label{eq:13scalars1}
	\nu = e^{(6\,\tilde{x}_{7}+3\,\tilde{x}_{8}+\sqrt{3}\,\tilde{x}_{9})/6}
				\begin{pmatrix}
					1 & \frac{x_{10}}{\sqrt{2}} & \frac{x_{11}}{\sqrt{2}} + \frac{x_{10}x_{12}}{4} \\
					0 & e^{-\tilde{x}_{8}} & \frac{e^{-\tilde{x}_{8}}\,x_{12}}{\sqrt{2}} \\
					0 & 0 & e^{-(\tilde{x}_{8}+\sqrt{3}\,\tilde{x}_{9})/2}
				\end{pmatrix},
\end{equation}
the matrices 
\begin{equation} \label{eq:13scalars2}
	\phi = 	\begin{pmatrix}
					0 & x_{4} & x_{5} \\
					-x_{4} & 0 & x_{6} \\
					-x_{5} & -x_{6} & 0
				\end{pmatrix}, \quad
	\xi = \begin{pmatrix}
				0 & 0 & 0 & x_{1} \\
				0 & 0 & 0 & x_{2} \\
				0 & 0 & 0 & x_{3}
			\end{pmatrix}, \quad
	\xi^{2} = \xi \xi^{T},
\end{equation}
and a dilaton $\tilde{\varphi} = \tilde{x}_{13}$. With this parametrization, the potential takes the form:
\begin{equation} \label{eq:scalarpotential}
	\begin{aligned}
		V & = 4\,e^{-4\tilde\varphi}+2\,e^{-2\tilde\varphi}\Big[-\tr\left(m+m^{-1}\right)+\tr\left(\phi m^{-1}\phi\right) -2\,\tr\left(\phi m^{-1}\xi^{2}\right)-2\,\tr\left(\xi^{2}\right)\\
		&\qquad\quad-\tr\left(\xi^{2}m^{-1}\xi^{2}\right)  +\frac{1}{2}\,\det\left(m^{-1}\right)\left(1-\tr\left(\phi^{2}\right)-\tr\left(\xi^{4}\right)+\tr\left(\xi^{2}\right)^{2}\right) \\
		&\qquad\quad +\frac{1}{2}\,{\rm T}\left(m^{-1}(\xi^{2}-\phi),(\xi^{2}+\phi)m^{-1},m+(\xi^{2}+\phi)m^{-1}(\xi^{2}-\phi)+2\,\xi^{2}\right)\\
		&\qquad\quad +\frac{1}{4}\,{\rm T}\left(m^{-1},m+(\xi^{2}+\phi)m^{-1}(\xi^{2}-\phi)+2\,\xi^{2},m+(\xi^{2}+\phi)m^{-1}(\xi^{2}-\phi)+2\,\xi^{2}\right)\Big],
	\end{aligned}
\end{equation}
where ${\rm T}\left(A,B,C\right)=\varepsilon_{mnp}\,\varepsilon_{qrs}\,A^{mq}B^{nr}C^{ps}$.
For later convenience, we define a thirteen-dimensional vector
\begin{equation} \label{eq:defvecX}
	\vec{X} = (x_1, x_{2}, x_{3}, x_{4}, x_{5}, x_6, \tilde{x}_7, \tilde{x}_8, \tilde{x}_9,x_{10}, x_{11}, x_{12}, \tilde{x}_{13}),
\end{equation}
encompassing all parameters. Note here that all dilaton fields are denoted with a tilde. This notation is adopted in anticipation of a later redefinition of the form \( x_i = e^{\tilde{x}_i} \) for these fields. 

\paragraph{}
The search for flat directions of the potential~\eqref{eq:scalarpotential} can then start form a study of its gradient $\nabla V$. However, carrying out the search for stationary points analytically is by far too complex, even reducing the number of variables or using a symbolic solver such as Mathematica~\cite{Mathematica}: the resulting expressions are too convoluted to be simplified into a manageable form, and do not yield usable relationships that express some variables in terms of others. This complexity does not, however, rule out the possibility that simpler solutions satisfying the condition of vanishing gradient may exist although the solver does not find them. 

In this work, we aim to identify such solutions. 
A preliminary numerical exploration suggests that attention can be focused on the five scalars $x_{1}, x_{2}, x_{4}, \tilde{x}_{8}$, and $x_{10}$. 
To ensure that the solutions we obtain in this truncation remain valid solutions of the complete theory, we first compute $\nabla V$ with all scalar fields included, \textit{i.e.} using eq.~\eqref{eq:scalarpotential}, and only then setting the remaining fields,
\begin{equation} \label{eq:defvecy}
	\vec{y} = (x_3,x_5,x_6,\tilde{x}_7,\tilde{x}_9,x_{11},x_{12},\tilde{x}_{13}),
\end{equation}
to zero, as detailed in sec.~\ref{sec:graddes}. By performing the differentiation before the truncation, we ensure that the resulting configurations satisfy the full equations of motion and are therefore legitimate solutions of the complete theory.

To identify the flat directions in the potential, we will combine numerical and symbolic tools. The procedure is as follows: first, we sample the underlying manifold by performing a gradient descent on a 5-d hypercube. The resulting cloud of points is subsequently analysed using local principal component analysis~(PCA) and clustering algorithms, which allow us to infer the dimension and topological structure of the manifold. After these properties are ascertained, we finally extract analytical constraints defining the manifold thanks to symbolic regression methods. In the next section, we introduce the symbolic regression algorithm, and defer the details of the numerical methods and results to the following one.

\section{Annealed Sequential Monte Carlo Sampler for Polynomial Symbolic Regression} \label{Sec:AIS-SMC}

	\subsection{General idea}
	The numerical sampling of the space of vacua (described in sec.~\ref{sec:graddes}) results in clouds of points embedded in a higher-dimensional space. 
	The aim of symbolic regression is to uncover interpretable mathematical expressions that best describe the embedding of these loci of solutions. 
	In the present context, the form of the supergravity potential \eqref{eq:scalarpotential} determining the solutions implies that the loci can be defined through polynomials $z_{m}$ such that $z_{m}\big(\vec{x}^{(i)}\big) = 0$, with $i \in \{1, \dots, n_{\rm points}\}$. Such polynomials will be called ``annihilating polynomials'' in the following.
	Finding annihilating polynomials involves navigating a vast, discrete, and often rugged search space of possible symbolic models, which poses significant challenges for traditional sampling methods. 
	Markov Chain Monte Carlo (MCMC) techniques such as Metropolis-Hastings or Gibbs sampling, while widely used, can struggle with poor mixing and often get trapped in local optima, especially in high-dimensional or multimodal spaces.

	To address these challenges, we formulate the search for annihilating polynomials as a determination of a probability density on the space of polynomials $E_{\rm pol}$ that stresses higher probabilities on polynomials~$z$ that minimise the error
	\begin{equation}
		\sum_{i=1}^{n_{\rm points}}z\big(\vec{x}^{(i)}\big)^{2}.
	\end{equation}
	Polynomials sampled from this density would thus be good approximations to annihilating polynomials. Such a probability density can take the form
	\begin{equation}
		\pi(z) \propto e^{-L(z)},
	\end{equation}
	with $L(z)$ a loss function chosen to be minimal on annihilating polynomials (its design will be discussed at the end of the sec.~\ref{sec:detailedAISSMC}). However, although one could in principle evaluate $\pi$ on a given polynomials, it is highly non trivial to draw a sample from it. Such a sample can be approximated using algorithms based on Sequential Important Sampling~\cite{Doucet2001,Liu2004} such as Sequential Monte Carlo samplers~\cite{del2006sequential}. They are inspired from Important Sampling (IS)~\cite{osti_4441935,RobertCasella2004}, a method to parametrise expectation values with respect to a target distribution $\pi$, that can be evaluated pointwise, but from which one does not know how to draw samples. IS uses another, typically simpler, distribution $\eta$ (called importance distribution) that one knows how to draw from (the easiest would be the distribution that assigns equal probability to all polynomials). It is based on the observation that the expectation value
	\begin{equation}
		\mathbb{E}_{\pi}[f] = \int_{E_{\rm pol}} f(z)\pi(z)\ \d z,
	\end{equation}
	for some function $f$ with respect to $\pi$, can be computed from the expectation $\mathbb{E}_{\eta}[w f]$, now computed with respect to the distribution $\eta$, with unnormalised weights $w = \pi/\eta$. Therefore, one can estimate $\mathbb{E}_{\pi}[f]$ from a Monte Carlo method: we draw a sample of polynomials $\{z^{(k)}\}_{k\in\llbracket 1, n_{\rm particles}\rrbracket}$ from $\eta$ and use the weighted average\footnote{Elements of this sample are often referred to as particles in the specialised literature~\cite{del2006sequential,DoucetTutorial,naesseth2024elementssequentialmontecarlo,Dai03072022}.}
	\begin{equation}
		\mathbb{E}_{\pi}[f] \simeq \displaystyle \sum_{k=1}^{n_{\rm particles}} w\big(z^{(k)}\big)\,f\big(z^{(k)}\big)\Bigg/\displaystyle \sum_{i=1}^{n_{\rm particles}} w\big(z^{(k)}\big).
	\end{equation}
	Stated differently, the target distribution $\pi$ can be approximated from a weighted collection of polynomials $\{z^{(k)},w^{(k)}\}_{k\in\llbracket 1, n_{\rm particles}\rrbracket}$ drawn from the importance distribution $\eta$ :
	\begin{equation}
		\pi(z) \simeq \displaystyle \sum_{k=1}^{n_{\rm particles}} w^{(k)}\,\delta\big(z-z^{(k)}\big)\Bigg/\displaystyle \sum_{k=1}^{n_{\rm particles}} w^{(k)},
	\end{equation}
	where $w^{(k)} = w\big(z^{(k)}\big)$. The sample admits $\pi$ as marginal distribution asymptotically when $n_{\rm particles}\rightarrow+\infty$. The accuracy of this estimation depends on the size of the sample and, crucially, on the choice of the importance distribution $\eta$, that needs to be close enough to $\pi$. In the following we will often use ``particles'' to denote the polynomials in the sample, as done in the IS literature.
	
	However, providing a good importance distribution $\eta$ remains a hard problem. In order to fix this point, we employ an Annealed Sequential Monte Carlo sampler (ASMC)~\cite{zimmermann2021nested,syed2024optimisedannealedsequentialmonte}, \textit{i.e.} a Sequential Monte Carlo sampler (SMC)~\cite{del2006sequential,DoucetTutorial,naesseth2024elementssequentialmontecarlo} enhanced by Annealed Importance Sampling (AIS)~\cite{neal1998annealedimportancesampling}. We give here a general overview of the method and defer the details to sec.~\ref{sec:detailedAISSMC}. SMC aims at estimating $\pi$ sequentially from multiple intermediate distributions $\pi_{n}$ that smoothly transition from an initial, tractable distribution $\pi_{0}$ (chosen to be easy to sample from) to the complex distribution $\pi$. Here we consider the sequence
	\begin{equation}
		\pi_{n}(z) = \pi_{0}(z)\,\frac{e^{-\beta_{n}\,L(z)}}{Z_{n}},
	\end{equation}
	with $\beta_{0}=0$, $(\beta_{n})$ tending towards $1$ and normalising constants $Z_{n}$. These distributions are approximated by weighted samples of particles $\big\{z^{(k)}_{n},w^{(k)}_{n}\big\}$ approximating $\pi_{n}$:
	\begin{equation} \label{eq:pifromweights}
		\pi_{n}(z) \simeq \sum_{k=1}^{n_{\rm particles}} \tilde{w}^{(k)}_{n}\,\delta(z-z^{(k)}_{n}),
	\end{equation}
	with  $\displaystyle \tilde{w}^{(k)}_n  = w^{(k)}_n/\sum_{k=1}^{n_{\rm particles}} w^{(k)}_{n}$ the normalised weights. The particles are incrementally deformed and reweighed to gradually sample the target distribution $\pi$. These deformations, for example a perturbation of the coefficients or a modification of a given monomial, are performed using a Markov kernel $q\big(z_{n}\vert z_{n-1}\big)$ that defines the probability to get $z_{n}$ when the current particle is $z_{n-1}$ (its definition will be discussed in sec.~\ref{sec:detailedAISSMC}). The weights are then updated:
	\begin{equation}
		w_{n}^{(k)} = w_{n-1}^{(k)}\,\alpha_{n}^{(k)},
	\end{equation}
	with the incremental importance weights $\alpha_{n}^{(k)}$ defined as
	\begin{equation}
			\alpha^{(k)}_{n} = \frac{\exp\Big(-\beta_{n+1}L\big(z_{n+1}^{(k)}\big)\Big)}{\exp\Big(-\beta_{n}L\big(z_{n}^{(k)}\big)\Big)}\,\frac{q\big(z^{(k)}_{n}\vert z^{(k)}_{n+1}\big)}{q\big(z^{(k)}_{n+1}\vert z^{(k)}_{n}\big)}.
	\end{equation}
	This process is guided by the temperature-like parameter $1/\beta_{n}$ that gradually emphasises the data likelihood, allowing for more efficient and tunable exploration of the polynomials landscape. The approximation procedure can be summed up as follows. We define the schedule of intermediate distributions $\pi_{n}$ using a temperature-like parameter, starting from an initial, easy to sample distribution $\pi_0$ which is parametrised by a set of particles and weights $\big\{z_0^{(k)},w_0^{(k)}\big\}$. We gradually mutate them with an annealing procedure to get a set of $\{z_n^{(k)},w_n^{(k)}\}$, which are used to parametrise $\pi_n$. As $n$ increases, the procedure converges such that the particles sample the target distribution $\pi$, \textit{i.e.} the polynomials $z_n^{(k)}$ tends to annihilating polynomials.
	
	This procedure has the disadvantage that the weights variance tends to increase, leading to weight degeneracy. ASMC includes a resampling procedure, that focusses computational effort on high-probability regions.  This combination of importance sampling, mutation and resampling maintains diversity among the particles and prevents premature convergence to suboptimal models. These features make ASMC particularly well-suited for symbolic regression tasks, where the search space is not only high-dimensional but also structured and discontinuous.

	\subsection{Detailed procedure}	\label{sec:detailedAISSMC}
	Let us now explain in more detail the procedure, based on ref.~\cite{del2006sequential,DoucetTutorial,naesseth2024elementssequentialmontecarlo,Dai03072022}. 
	The goal is to obtain particles sampled from a probability distribution $\pi(z)$, where $z$ runs over a space of polynomials $E_{\rm pol}$, that assigns high probabilities to those polynomials that annihilate the data, and therefore will be used to identify their symbolic expressions. We restrict ourselves to polynomials of maximum degree ${\rm max}_{\rm degree}$. Given the number of variables $n_{\rm var}$, this fixes the number of different possible monomials to $\binom{{\rm max}_{\rm degree}+n_{\rm var}}{{\rm max}_{\rm degree}}$. We further restrict $E_{\rm pol}$ to include only polynomials with a maximum number of ${\rm max}_{\rm mon}$ monomials. The space $E_{\rm pol}$ is then finite dimensional.

	We reconstruct the density function $\pi$ by series of density functions $\pi_n(z)$, $n\in\llbracket0,n_{\rm epochs}\rrbracket$, such that $\pi_n\to \pi$ as $n\to n_{\rm epochs}$. On every annealing step, $\pi_n$ is defined in terms of an unnormalised density $\gamma_n$ and a normalisation constant as $\pi_n(z) = \gamma_n(z)/Z_n$. The unnormalised density at level $n$ is given in terms of a prior distribution $\pi_0(z)$ over the space $E_{\rm pol}$ of polynomials and a loss function $L(z)$, as
		\begin{equation}\label{eq:gamman}
			\gamma_n(z) = \pi_0(z) \: \exp \Big( -\beta_n L(z)\Big),
		\end{equation}
	where the inverse temperature constants $\beta_n$ are taken to evolve monotonically as $\beta_0 = 0 < \beta_1 <\dots< \beta_{n_{\rm epochs}}$. The importance given to each polynomial $z$ is thus updated at each step, with polynomials~$z$ with high loss being disfavoured. This update is controlled by the inverse temperature constants, enabling deeper exploration during the early stages and protecting the procedure from being trapped in local minima. We refer to eq.~\eqref{eq:lossreg} for the definition of the loss function $L(z)$. We choose the prior distribution $\pi_{0}$ to be flat on the space of polynomials $E_{\rm pol}$, giving equal importance to all of them, which makes easy the initial sampling of the weighted particles $\big\{z^{(k)}_{0},w^{(k)}_{0} = 1/n_{\rm particles}\big\}$.\footnote{Note that with this choice there is a higher probability to draw polynomials with multiple monomials.} The random draw is done on the monomials of each polynomials, and the coefficients are chosen randomly in the range $[-2,2]$.

	At each epoch $n\geq1$, we do not directly approximate $\pi_{n}$, but rather a distribution $\tilde{\pi}_{n}\big(z_{0:n}\big) = \tilde{\gamma}_{n}\big(z_{0:n}\big)/\tilde{Z}_{n}$ defined on $E_{\rm pol}^{n+1}$, with $z_{0:n} = \big(z_{0},\ldots,z_{n}\big)$, that describes the global trajectory of the particles until step $n$. $\tilde{\gamma}_{n}$ is defined thanks to a backward propagation kernel $K\big(z_{i-1}\vert z_{i}\big)$:
	\begin{equation}
		\tilde{\gamma}_{n}\big(z_{0:n}\big) = \gamma_{n}\big(z_{n}\big)\,\prod_{i=1}^{n} K\big(z_{i-1}\vert z_{i}\big).
	\end{equation}
	This way, the distribution $\tilde{\pi}_{n}$ describes $\pi_{n}$ as well as the path that followed the particles from epochs $0$ to~$n$. $\tilde{\pi}_{n}$ evaluates not only the probability that the particle is in a given state $z_{n}$, but also the probability that the particle has followed the path from $z_{0}$ to $z_{n}$. This allows for more accuracy, correcting potential bias introduced by the mutations, and results in a setup that is more tunable.
	
	At the beginning of epoch $n\geq1$, $\tilde{\pi}_{n-1}$ is approximated \textit{via} Importance Sampling from an importance distribution $\eta_{n-1}$, \textit{i.e.} from a set of particles and weights $\big\{z_{n-1}^{(k)},\ w_{n-1}^{(k)}\big\}$ using eq.~\eqref{eq:pifromweights}, with $\big\{z_{n-1}^{(k)}\big\}$ drawn from $\eta_{n-1}$ and $w_{n-1}^{(k)} = \tilde{\gamma}_{n-1}\big(z_{0:n-1}^{(k)}\big)/\eta_{n-1}\big(z_{0:n-1}^{(k)}\big)$.\footnote{We define $\eta_{0} = \pi_{0} = \tilde{\pi}_{0}$.} The importance density $\eta_{n-1}$ gets mutated using a forward propagation Markov kernel $q\big(z_{n}\vert z_{n-1}\big)$, defining the probability to get $z_{n}$ when the current state is $z_{n-1}$:
	\begin{equation}
		\eta_{n}\big(z_{0:n}\big) = \eta_{n-1}\big(z_{0:n-1}\big)\,q\big(z_{n}\vert z_{n-1}\big).
	\end{equation}
	We then use Importance Sampling to approximate $\tilde{\pi}_{n}$: we draw particles $\big\{z^{(k)}_{n}\big\}$ from $\eta_{n}$, and compute the weights
	\begin{equation} \label{eq:weightsupdate}
		w_{n}^{(k)} = \frac{\tilde{\gamma}_{n}\big(z_{0:n}^{(k)}\big)}{\eta_{n}\big(z_{0:n}^{(k)}\big)} = w_{n-1}^{(k)}\,\alpha_{n}^{(k)},
	\end{equation}
	with the incremental importance weights
	\begin{equation}	\label{eq: incremental_importance}
			\alpha^{(k)}_{n} =  \frac{\gamma_{n}\big(z_{n}^{(k)}\big)\,K\big(z^{(k)}_{n-1}\vert z^{(k)}_{n}\big)}{\gamma_{n-1}\big(z_{n-1}^{(k)}\big)\,q\big(z^{(k)}_{n}\vert z^{(k)}_{n-1}\big)} = \frac{\exp\Big(-\beta_{n+1}L\big(z_{n+1}^{(k)}\big)\Big)}{\exp\Big(-\beta_{n}L\big(z_{n}^{(k)}\big)\Big)}\,\frac{K\big(z^{(k)}_{n}\vert z^{(k)}_{n+1}\big)}{q\big(z^{(k)}_{n+1}\vert z^{(k)}_{n}\big)}.
	\end{equation}

	For the implementation of the forward propagation kernel $q\big(z_{n}\vert z_{n-1}\big)$, we perform an AIS-style MCMC algorithm. On every epoch, we make some move in the space of polynomials, and then accept or reject those new polynomials based on a fixed rate. The moves that have been allowed in our implementation are the following.
	\begin{itemize}
		\item \textbf{Coefficient perturbation}\quad Given a polynomial, we choose one of its coefficients at random and modify it by a Gaussian noise distributed as $\mathcal{N}(0,\sigma^2)$, \textit{e.g.}
		\begin{equation} \label{eq:coeffperturb}
			2\, x_1x_{2} + 3\, x_2^2 \longmapsto  2.1\, x_1x_{2} + 3\, x_2^2.
		\end{equation}
		\item \textbf{Variable multiplication}\quad Given a polynomial, we pick randomly one of its monomials, and multiply it by one of the available variables, \textit{e.g.}
		\begin{equation}
			2\, x_1x_{2} + 3\, x_2^2 \longmapsto  2\, x_1 x_2^{2} + 3\, x_2^2.
		\end{equation}
		\item \textbf{Variable division}\quad Given a polynomial, we choose randomly one of its monomials, and divide it by one of its variables, \textit{e.g.}
		\begin{equation}
			2\,x_1 x_2 + 3\, x_2^2 \longmapsto  2\, x_1 + 3\, x_2^2.
		\end{equation}
	\end{itemize}
	Each of these operations is chosen randomly, with probabilities $p_{\rm shift}$, $p_{\rm multiply}$, and $p_{\rm divide}$.
	A common choice of backward kernel in such context is the reversed forward kernel $K\big(z_{n-1}\vert z_{n}\big) = q\big(z_{n-1}\vert z_{n}\big)$.
	After performing these updates, the change is accepted based on the temperature-dependent rate
	\begin{equation}\label{eq:acceptanceratio}
		A\big(z_{n}^{(k)},z_{n-1}^{(k)}\big) = \mathrm{min} \left(1,\alpha_{n}^{(k)}\right) = \mathrm{min} \left(1,\frac{\gamma_{n}\big(z_{n}^{(k)}\big)\,q\big(z^{(k)}_{n-1}\vert z^{(k)}_{n}\big)}{\gamma_{n-1}\big(z_{n-1}^{(k)}\big)\,q\big(z^{(k)}_{n}\vert z^{(k)}_{n-1}\big)}  \right).
	\end{equation}
	We then draw a number $u\sim \mathrm{Uniform}(0,1)$, accept the new particle if $u<A\big(z_{n+1}^{(k)},z_{n}^{(k)}\big)$ and reject it otherwise. This way we accept systematically all moves that increase the weights, and some of those that lower the weights, allowing for some exploration of the space of polynomials.	Once the particles have been mutated, the evolution of the weights is computed using eq.~\eqref{eq:weightsupdate}.

	To ensure an efficient exploration of the landscape, we would like to avoid having too many particles with low weights: as the algorithm progresses, particles with negligible weights contribute minimally to the approximation of the target distribution (see eq.~\eqref{eq:pifromweights}), leading to a concentration of the effective sample on a subset of particles with substantial weights. This can be estimated using the effective sample size~\cite{ca54d442-d5bd-32ea-a920-38aa7d8a4043}
	\begin{equation}
		{\rm ESS}_{n} = \Bigg(\sum_{k=1}^{n_{\rm particles}} \tilde{w}^{(k)}_{n}{}^{2}\Bigg)^{-1},
	\end{equation}
	where the $\tilde{w}^{(k)}_{n}$ are normalised weights deduced from the $w^{(k)}_{n}$. ${\rm ESS}_{n}$ takes values between $1$ (complete degeneracy, there is only one meaningful particle) and $n_{\rm particles}$ (no degeneracy, all particles contribute equally). We resample the particles when the ESS falls below a predetermined threshold, typically $n_{\rm particles}/2$. During resampling, new particles are drawn from the empirical distribution defined by the current particle weights $\tilde{w}^{(k)}_{n}$.	This procedure results in a new particle set where high-weight particles may appear multiple times, while low-weight particles may be eliminated entirely. Following resampling, all particle weights are reset to uniform values $w_{n}^{(k)} = 1/n_{\rm particles}$, as the information previously encoded in the weight distribution has been incorporated into the spatial distribution of the resampled particles. This resampling mechanism ensures that computational resources are concentrated on exploring the most promising regions of the state space.

	\begin{figure}[h!]
			\centering
			\begin{minipage}{15cm}
				\begin{tcolorbox}[
					colback=white,
					colframe=myblue,
					title=\textbf{ASMC Algorithm},
					fonttitle=\sffamily\large,
					halign title=flush center,
					boxrule=2pt,
					left=6pt,
					right=6pt,
					top=8pt,
					bottom=8pt
					]
					\begin{tikzpicture}[node distance=1.6cm and 0.8cm, scale=0.85, transform shape]
						
						\node (start) [startstop] {Start};
						
						\node (init_particles) [process, below of=start] {Initialize particles \& weights};
						
						\node (compute_beta) [process, below of=init_particles] {Compute/adapt inverse temperature $\beta$};
						
						\node (weight_update) [process, below of=compute_beta] {Compute losses and update weights};
						
						\node (ess_check) [decision, below of=weight_update] {Effective Sample Size $<$ Threshold?};
						
						\node (resample) [process, right=3.8cm of ess_check] {Resample particles};
						
						\node (mutation) [process, below of=ess_check, yshift=-0.4cm] {Mutate particles with MCMC kernel};
						
						\node (acceptance) [process, below of=mutation] {Accept/Reject particles};
						
						\node (end_iter) [decision, below of=acceptance] {Last step?};
						
						\node (exploit) [process, below of=end_iter, yshift=-0.4cm] {Optional local search};
						
						\node (output) [process, below of=exploit] {Return particles};
						
						\node (stop) [startstop, below of=output] {End};
						
						\draw [arrow] (start) -- (init_particles);
						\draw [arrow] (init_particles) -- (compute_beta);
						\draw [arrow] (compute_beta) -- (weight_update);
						\draw [arrow] (weight_update) -- (ess_check);
						
						\draw [arrow] (ess_check) -- node[above, arrow_label] {Yes} (resample);
						\draw [arrow] (resample) |- (mutation);
						
						\draw [arrow] (ess_check) -- node[left, arrow_label] {No} (mutation);
						\draw [arrow] (mutation) -- (acceptance);
						\draw [arrow] (acceptance) -- (end_iter);
						
						\draw [arrow] (end_iter) -| node[pos=0.25, below, arrow_label] {No} ++(-4.5,0) |- (compute_beta);
						
						\draw [arrow] (end_iter) -- node[right, arrow_label] {Yes} (exploit);
						\draw [arrow] (exploit) -- (output);
						\draw [arrow] (output) -- (stop);
						
					\end{tikzpicture}
				\end{tcolorbox}
			\end{minipage}
			\caption{Flow chart of the Annealed Sequential Monte Carlo sampler algorithm.}
			\label{ASMCflowchart}
		\end{figure}
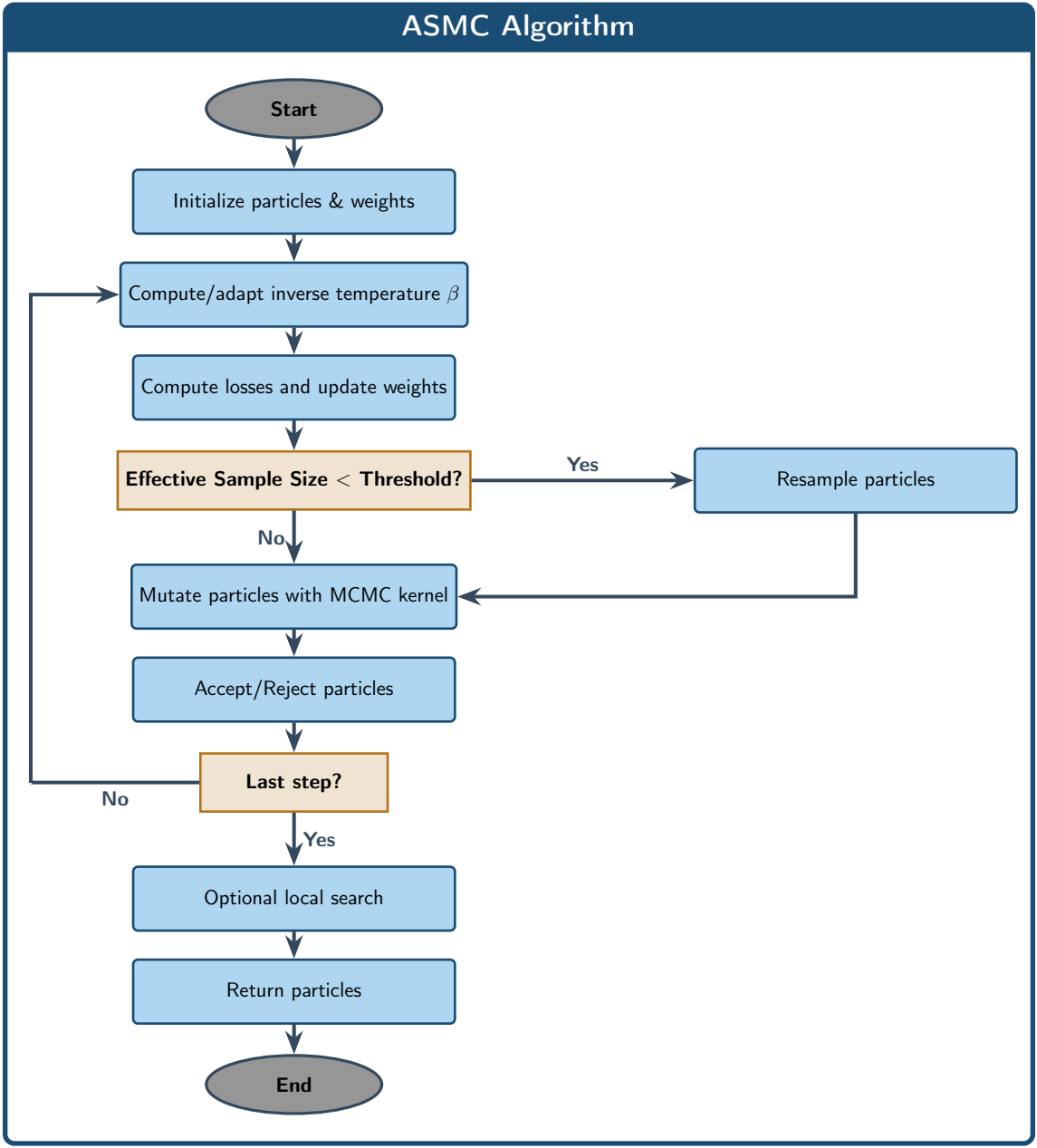

	To sum up, the algorithm starts initially with $n_{\rm particles}$ particles $\big\{z^{(k)}_{0}\big\}$ sampled from the prior distribution $\eta_{0}=\pi_{0}$, with equal weights $\big\{w^{(k)}_{0} = 1/n_{\rm particles}\big\}$. At each epoch $n\geq1$, the intermediate target distribution $\tilde{\pi}_{n}$ is estimated from the sample of weighted particles $\big\{z^{(k)}_{n}, w^{(k)}_{n}\big\}$, obtained from propagation using MCMC moves, reweighing~\eqref{eq:weightsupdate} and, if needed, resampling. At the final epoch $n_{\rm epochs}$, the particles $\big\{z^{(k)}_{n_{\rm epochs}}, w^{(k)}_{n_{\rm epochs}}\big\}$ approximate a sampling of the target distribution $\pi$. A flow chart of the full ASMC procedure is given in fig.~\ref{ASMCflowchart}. Note that in practice the reweighing and resampling of particles $z_{n-1}^{(k)}$ are done at the beginning of the epoch $n$.

	\paragraph{}
	To compute the acceptance ratio and weights \textit{via} eq.~\eqref{eq:acceptanceratio} and \eqref{eq: incremental_importance} and we need to consider the loss function in eq~\eqref{eq:gamman}.
	In the present context, we use the following loss, where the polynomials have been parameterised as $z = \sum_m c_m X_m$ with $X_m$ denoting the possible monomials up to a given degree,%
	\footnote{So, if we have $x$ and $y$ as variables, and the maximum degree is 2, then $X_m$ are $1, x, y, x^2, y^2, x y$.}
	\begin{equation} \label{eq:lossreg}
		L(z) = \sum_{i=1}^{n_{\rm points}} z(x^{(i)})^2 + \frac{\lambda}{\displaystyle \sum_m |c_m|}\,.
	\end{equation}
	Here, the first term is just the sum of the square of the polynomial evaluated on the data, and is therefore minimal when the polynomial annihilates the data. 
	The second term is a regularisation factor that prevents the algorithm to send all the coefficient to 0, which would give a trivial solution to the problem. 
	We typically take $\lambda \sim \mathcal{O}(10^3)$. 
	Together with the loss function, we also need to specify a cooling schedule for $\beta$. We will discuss different approaches in the next section.
	
	\paragraph{}
	Once the ASMC loop is over, we end up with a total of $n_{\rm particles}$ polynomials, which in principle should be close to annihilate our data, but whose coefficients may need some refinement. To deal with it, for each polynomial we run a fine tuning loop as described in sec.~\ref{sec:1runanalysis}.
	
\section{Numerical Analysis and Results} \label{sec:results}
After describing the theory we want to address in sec.~\ref{sec:sugra}, and the symbolic regression method to be employed in sec.~\ref{Sec:AIS-SMC}, we now proceed to detail the numerical analysis and present our main results. 
We start by describing how we sampled the space of solutions using gradient descent. Once the manifold correctly sampled, we determine its dimension using a local principal component analysis and a clustering procedure. We finally exploit the ASMC algorithm described in sec.~\ref{Sec:AIS-SMC} to find an analytical characterisation of the manifold.

	\subsection{Sampling the manifold: gradient descent} \label{sec:graddes}
	To perform the gradient descent, we initialise randomly and uniformly points within a hypercube of range $[-2,2]$. This choice is important to ensure that points are not restricted to the inner range~$[-1,1]$, which will be crucial for the symbolic regression.\footnote{The rationale is that our symbolic regression evaluates the data points $\vec{x}^{(i)}$ using a candidate polynomial function~$z$. When all data points lie within the interval $[-1, 1]$, the algorithm tends to favor high-degree polynomials artificially lowering the loss. Introducing data points with absolute values greater than 1 mitigates this bias.} As the example we are focussing on has a five-dimensional parameter space, we choose to generate $n_{\rm points}=10^5$ points, since we aim at populating all five directions and want approximately $\mathcal{O}(10)$ points per direction. As the solutions lie within the five dimensional space, its intrinsic dimension is less than or equal to 5. The value $10^5$ then serves as a conservative upper bound for the number of points needed to adequately sample the manifold.
	
	We then perform a gradient descent on the points using TensorFlow’s automatic differentiation \cite{tensorflow2015-whitepaper}. The loss function is defined as
	\begin{equation} \label{eq:lossgraddes}
		\mathcal{L} = \sum_{i = 1}^{n_{\rm points}} \left|\left|\left.\nabla V(\vec{X}^{(i)})\,\right|_{\vec{y}^{\,(i)}=0}\right|\right|^2,
	\end{equation}
with the vectors $\vec{X}$ and $\vec{y}$ given in~\eqref{eq:defvecX} and~\eqref{eq:defvecy}. The exponents $(i)$ denote the data points, with $i\in\llbracket1,n_{\rm points}\rrbracket$. This way we only keep $\vec{x}^{(i)} = (x^{(i)}_1,x^{(i)}_2,x^{(i)}_4,\tilde{x}^{(i)}_8,x^{(i)}_{10})$ alive after we have taken the derivative. As already mentioned above, we use the analytic formula for $\nabla V$, but we could have performed the gradient descent by fully using automatic differentiation.

	\begin{figure}[b!]
		\centering
		\includegraphics[scale = 0.75]{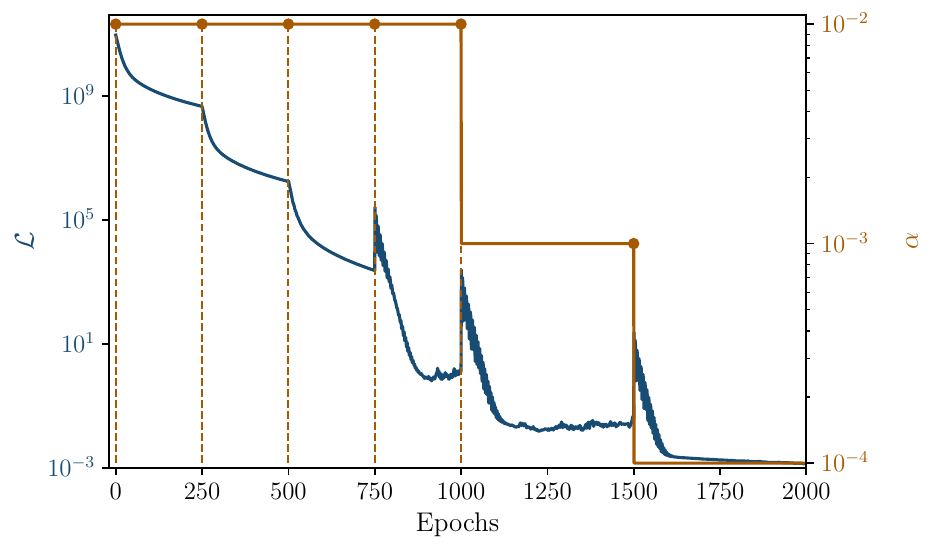}
		\caption{Evolution of the loss function $\mathcal{L}$ in~\eqref{eq:lossgraddes} during gradient descent (right axis, in blue) and corresponding learning rate schedule $\alpha$ (left axis in orange), both plotted in logarithmic scale. Dashed lines indicate epochs at which the optimizer was reinitialized.}
		\label{fig:loss_grad_des}
	\end{figure}

	The Adam optimizer was employed throughout the gradient descent procedure~\cite{kingma2017adammethodstochasticoptimization}. As an accelerator of convergence, we observed that periodically reinitialising the optimizer significantly improved the convergence rate. It was reinitialised at iterations {250, 500, 750}, with the learning rate $\alpha$ fixed at $10^{-2}$. At iteration 1000, the optimizer was reinitialised once more, this time with a reduced learning rate of~$10^{-3}$. A final reinitialisation was performed at iteration 1500, setting the learning rate to $10^{-4}$, and the optimization was continued for an additional 500 epochs. The evolution of the loss function~\eqref{eq:lossgraddes}, along with the learning rate schedule, is shown in fig.~\ref{fig:loss_grad_des}. As can be seen in this figure, the convergence rate improves significantly each time the optimizer is rebooted. We also observe that, following the last few reinitialisations, the loss exhibits a small bump immediately after the restart. We interpret this behaviour as follows: the learning rate gets internally adjusted by the Adam optimizer during the descent, and may well be smaller than the instructions when the reset occurs. The learning rate then gets suddenly increased, and some points that previously had low loss values may momentarily worsen before benefiting from faster convergence. This effect of faster convergence after the reinitialization is likely due to points initially located in regions with weak attractive basins being pushed toward areas where the potential gradient is steeper, thus accelerating their convergence. This strategy introduces the risk of desampling certain regions in favour of others, introducing the risk to hide some flat directions in the potential. However, we believe that with enough points, it is very unlikely for a flat direction to be completely desampled. This method bears similarity to the concept of warm restarts introduced in ref.~\cite{loshchilov2017sgdrstochasticgradientdescent}, where the authors reset the learning rate to some value at each reset, without reinitialising the whole optimizer. However, we found that simply scheduling the learning rate without resetting the optimizer yielded slower convergence. We interpret this as follows: when the optimizer is reinitialised, it effectively "forgets" its past gradient history. As a result, the actual learning rate used corresponds more closely to the specified value, rather than being internally adjusted based on accumulated past gradients. This effect seems to contribute to faster convergence in our case.
	
	\begin{figure}
		\centering
		\subfigure{ \label{fig:log_grad_V}
			\begin{tikzpicture}
		  		\draw (0,0) node (fig1) {\includegraphics[scale=0.7]{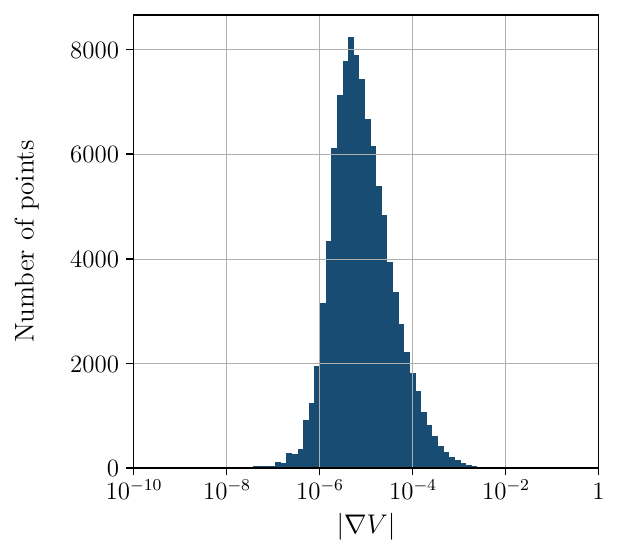}};
		  		\draw ($(fig1.north west) + (0,0)$) node {\small (a)};
		  		\draw (9,0) node (fig2) {\includegraphics[scale=0.7]{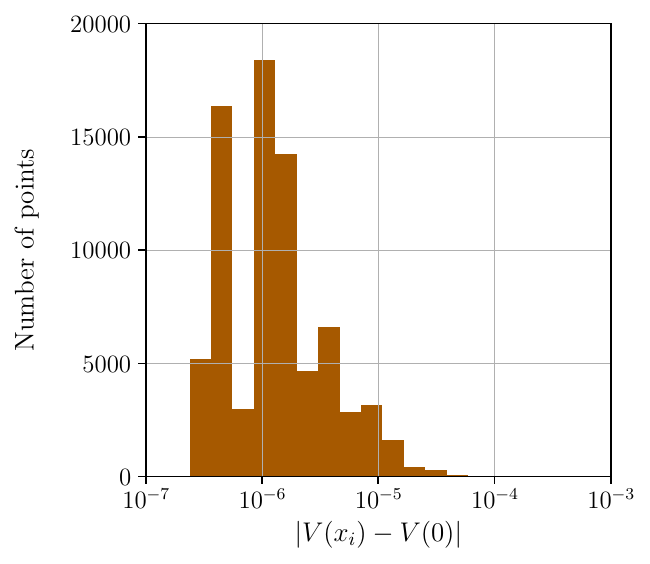}};
		  		\draw ($(fig2.north west) + (0,0)$) node {\small (b)};
		  	\end{tikzpicture}
		}
		\subfigure{\label{fig:log_Vp4}}
		\vspace{-1cm}
		\caption{(a) Histogram of the norm of the gradient for each point ($x$-axis is in $\log$ scale). (b)~Histogram of the distance of the value of the potential to $V(0)$ for each point ($x$-axis is in $\log$ scale). 
		}
		\label{fig:Analyze_V_points}
	\end{figure}
	
	Upon completion, the loss function converges around $10^{-3}$. The distribution of the values of the gradient at each of the data points is plotted in fig.~\ref{fig:log_grad_V}. More than $93\%$ of the points converged to values of $\vert\nabla V\vert$ lower than $10^{-4}$, and more than $99\%$ of them exhibited gradient norms smaller than~$10^{-3}$, yielding a satisfying sampling of the flat directions. As can be observed in fig.~\ref{fig:log_Vp4}, all data points converged to values close to $V(0)=-4$. Specifically, nearly all points attained $V(0)$ within an absolute error of at least $10^{-4}$, with the exception of four points whose deviations were of the order of $10^{-3}$ (not visible on the graph). As the origin $\vec{X}=0$ of the parameter space corresponds to the three-dimensional truncation of the round ${\rm AdS}_{3}\times S^{3}$ solution of half-maximal supergravity in six dimensions with $V(0)=-4$, this confirms that the data points lie within flat directions of the potential.

	\begin{figure}[t!]
		\centering
		\makebox[\textwidth][c]{\includegraphics[scale = 0.7]{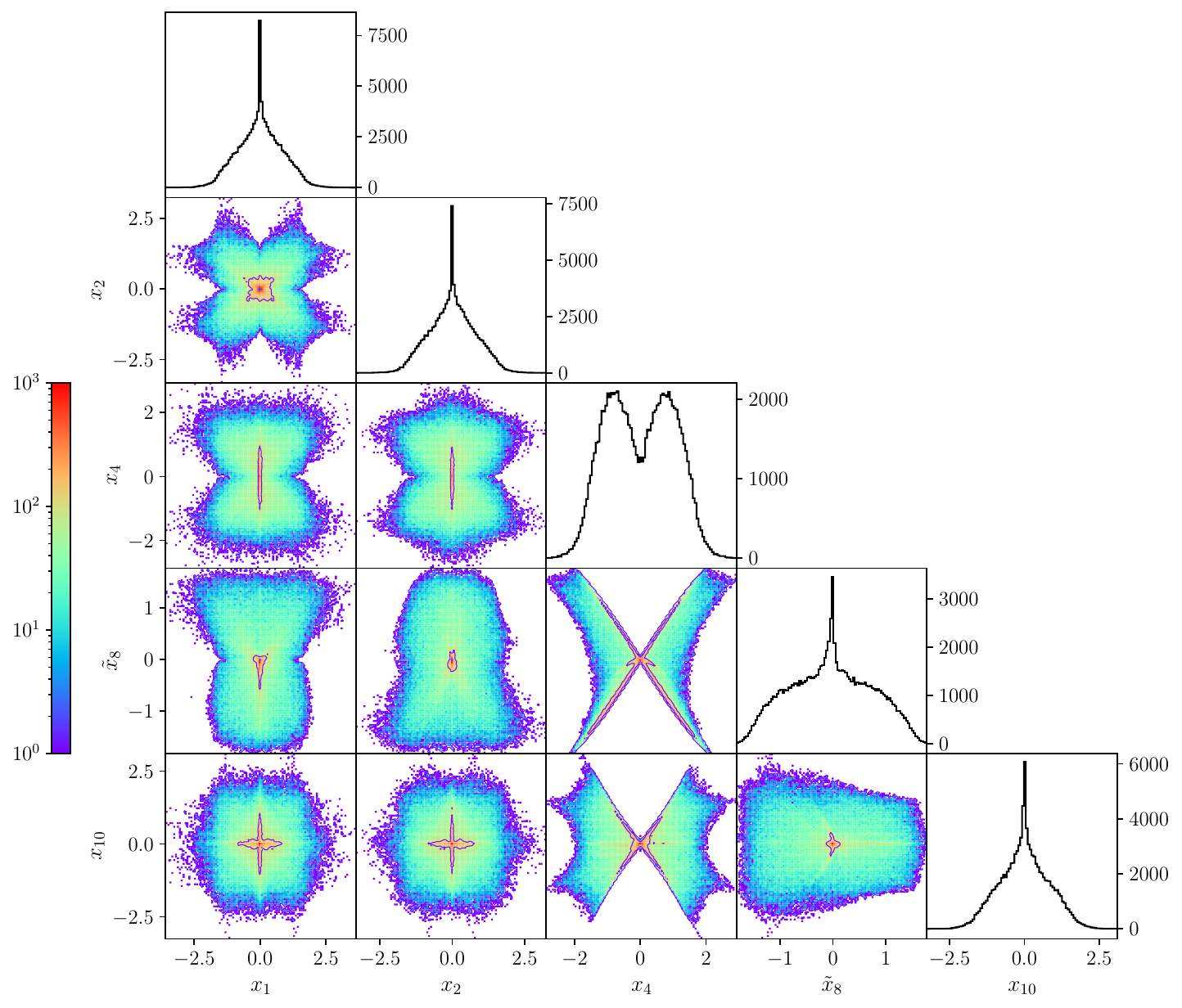}}
		\caption{Triangular plot showing all $2d$ projections and $1d$ histograms of the data after the gradient descent.}
		\vspace{-0.5cm}
		\label{triangular_plot_1_2_4_8_10}
	\end{figure}
	
	As a first visualisation, we present a tomography of the data in fig.~\ref{triangular_plot_1_2_4_8_10}. This figure shows all possible 2d projections of the data, along with the 1d histograms of each coordinate after gradient descent. Note that all directions appear to be well populated. There also seem to be non-trivial correlations in the data, see for example the $x_1/x_2$, $x_4/\tilde{x}_8$ or $x_4/x_{10}$ graphs. They could be genuine correlations, or may result from larger basins of attraction. Thanks to the analysis of sec.~\ref{sec:localanalysis} and~\ref{sec:AIS-SMC_application}, we find that the structures in the $x_1/x_2$ plot are artefacts of the gradient descent, whereas the ones seen in the $x_4/\tilde{x}_8$ or $x_4/x_{10}$ plots reflect genuine features of the manifold. 

	\subsection{Local analysis: extracting the dimension} \label{sec:localanalysis}
\begin{figure}
		\centering
		\includegraphics[scale=0.75]{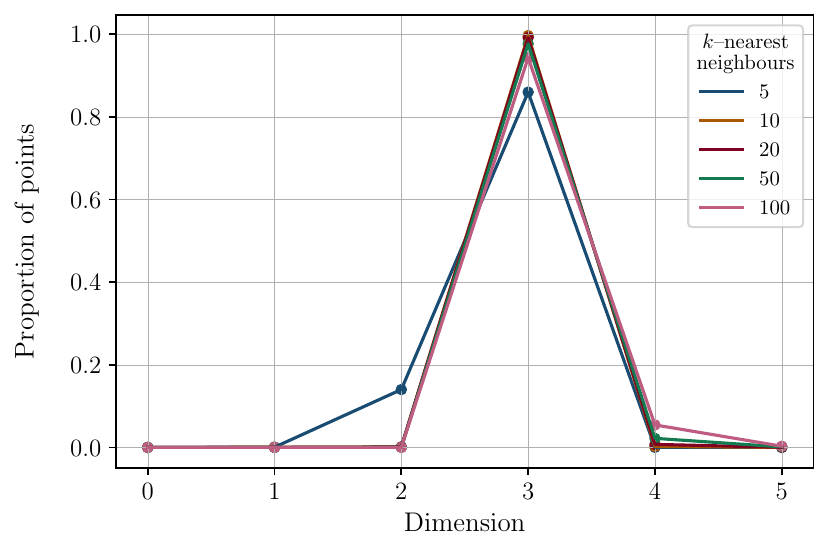}
		\caption{Results of the local PCA analysis. The $x$-axis shows the dimension inferred by the algorithm, and the $y$-axis indicates the proportion of points for which that dimension was found. Each curve corresponds to a different value of $k$ in the $k$-nearest neighbours.}
		\label{fig:local_pca}
	\end{figure}

	Once the gradient descent has been completed and the flat directions sampled, the next step is to identify the structure of the underlying manifold. Our goal is to eventually obtain an analytical expression, not just a numerical description. Before applying symbolic regression to search for such an expression, we first perform some exploratory analyses to better understand the data. Specifically, we aim at determining the dimension of the manifold and whether it consists of a single connected component or multiple disjoint components (\textit{e.g.} two intersecting hyperplanes). To this end, we apply a local Principal Component Analysis (PCA).\footnote{Principal Component Analysis is a dimensionality reduction technique that transforms high-dimensional data into a lower-dimensional space while preserving the maximum amount of variance, see for example ref.~\cite{scikit-learn}.} 
	For each point, we identify its $k$ nearest neighbours and perform a PCA on that local neighbourhood. This procedure allows us to determine how many principal directions are needed to explain a given proportion $\epsilon$ of the data variance. In other words, it provides an estimate of the local dimensionality around each point, \textit{i.e.} the dimension of its local tangent space. We perform this analysis for several values of $k$, namely $k \in \{5, 10, 20, 50, 100\}$, and we fix $\epsilon = 0.99$. The results are presented in fig.~\ref{fig:local_pca}. We observe that for every choice of $k$, there is a prominent peak at $d = 3$, suggesting that the underlying manifold is three-dimensional. For $k = 5$, a noticeable fraction of points are assigned dimension 2. This can be attributed to the fact that if the genuine dimension is 3, then selecting only 5 neighbours may not sufficiently populate all three directions, leading the algorithm to underestimate the dimensionality for a fraction of the points. Additionally, for $k \geq 20$, we observe an increasing number of points being assigned dimensions 4 or 5. This behaviour can be explained by the loss of locality when the number of neighbours becomes too large: increasing $k$ results in a coarser approximation, and the algorithm may then incorporate points that are no longer truly local. This artificial enlargement of the neighbourhood can cause the estimated local dimensionality to rise. We thus infer that the manifold under investigation has an intrinsic dimension of 3.

\begin{figure}[t!]
		\centering
		\centerline{
			\subfigure{ \label{fig:3dx4x8x10}
				\begin{tikzpicture}
					\draw (0,0) node (fig11) {\includegraphics[scale=0.7, trim = 0.25cm 0.2cm 0.25cm 1.2cm, clip]{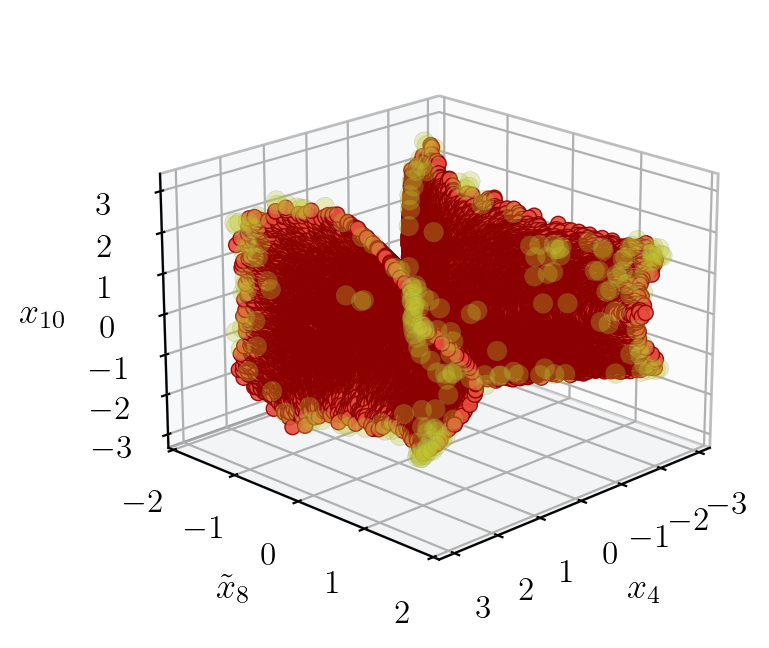}};
			  		\draw ($(fig11.north west) + (0,0)$) node {\small (a)};
			  		\draw (5.75,-0.25) node (fig12) {\includegraphics[scale=0.7, trim = 0.5cm 0.1cm 1.3cm 1.4cm, clip]{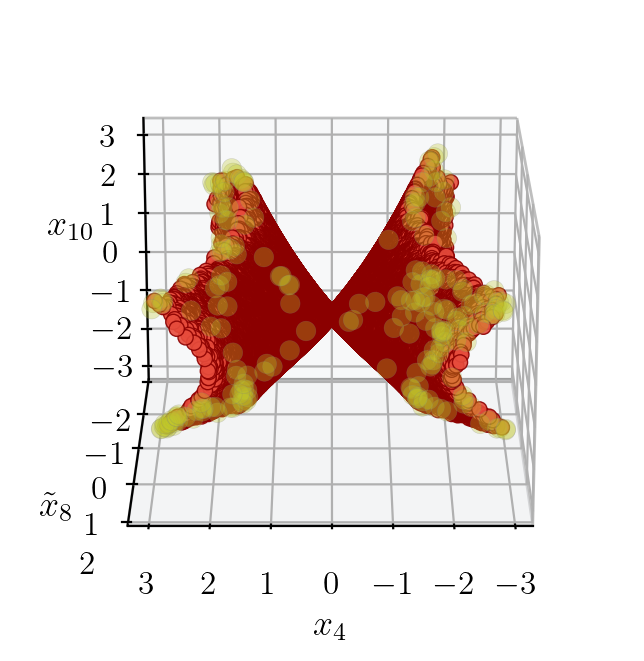}};
			  		\draw (11.5,0) node (fig13) {\includegraphics[scale=0.7, trim = 0.25cm 0.2cm 0.2cm 1.2cm, clip]{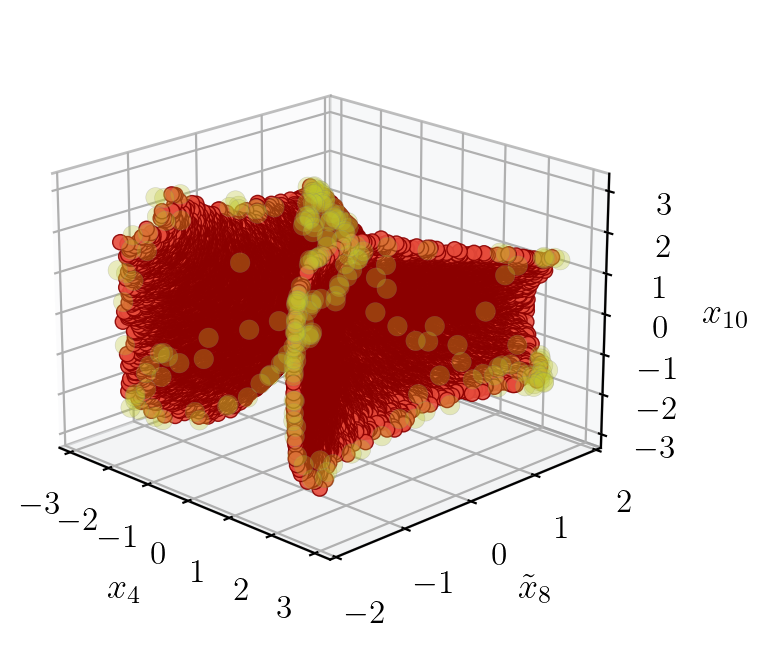}};

			  		\begin{scope}[shift={(0,-5.75)}]
			  			\draw (-0.25,0) node (fig21) {\includegraphics[scale=0.7, trim = 0.25cm 0.2cm 0.25cm 1.2cm, clip]{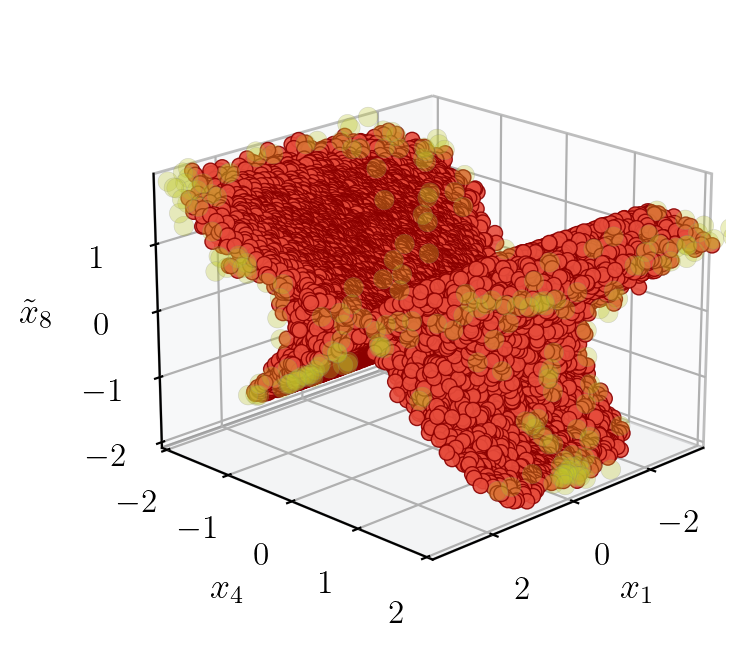}};
				  		\draw ($(fig11.north west) + (0,-5.5)$) node {\small (b)};
				  		\draw (5.75,-0.25) node (fig22) {\includegraphics[scale=0.7, trim = 0.2cm 0.1cm 0.35cm 1.4cm, clip]{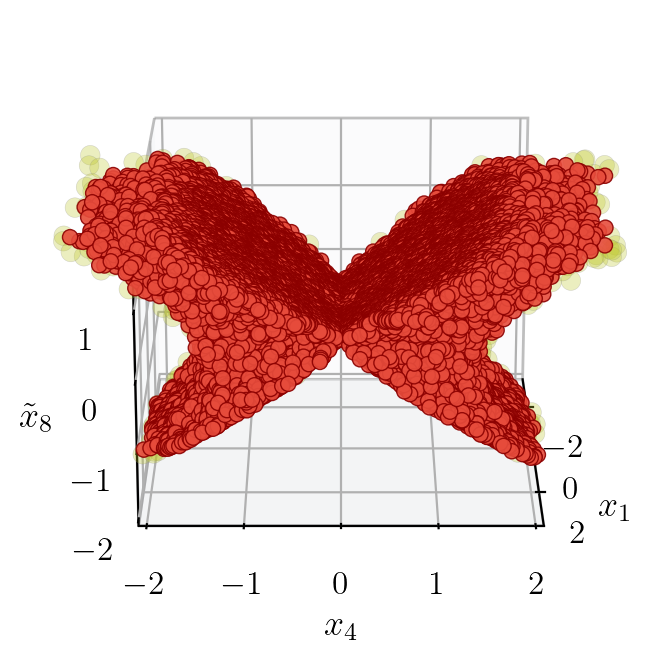}};
				  		\draw (11.75,0) node (fig33) {\includegraphics[scale=0.7, trim = 0.6cm 0.2cm 0.2cm 1.2cm, clip]{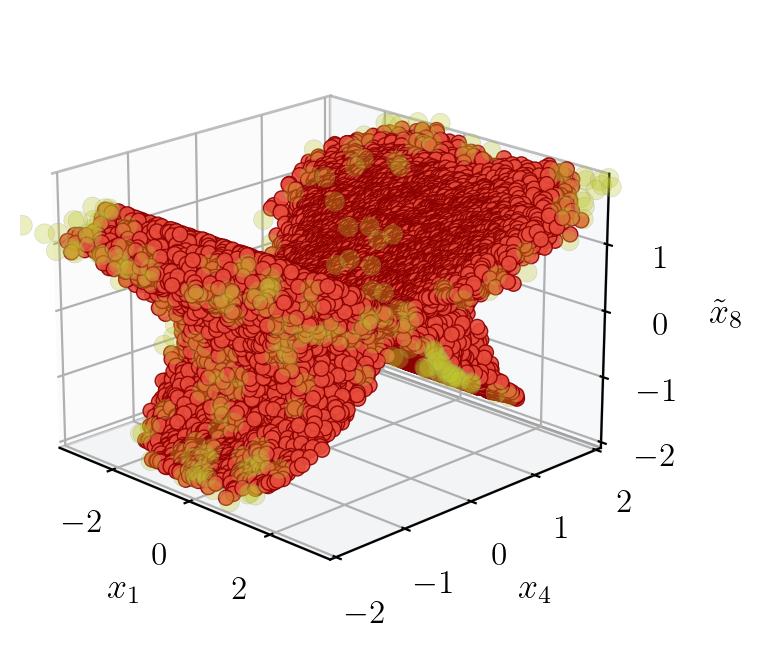}};
			  		\end{scope}

			  		\draw [thick, clusteringred1, fill=clusteringred2,opacity=0.9] (12,-2.6) circle [radius=2.5pt] node [right=0.3cm, black, opacity=1] {\small Cluster};
			  		\draw [gray, fill=clusteringyellow1, opacity=0.6] (12,-3.1) circle [radius=3pt] node [right=0.3cm, black, opacity=1] {\small Not assigned};
			  		\draw [rounded corners=2pt] (11.75,-2.35) rectangle +(2.75,-1.05);
			  	\end{tikzpicture}
			}
		}
		\subfigure{\label{fig:3dx1x4x8}}
		\vspace{-1cm}
		\caption{3d projections of the data set in selected coordinates: (a) $(x_4, \tilde{x}_8, x_{10})$ and (b) $(x_1, x_4, \tilde{x}_8)$. The colors are assigned by the clustering algorithm: the red points correspond to the cluster, while the yellow points were not assigned to any cluster. The size of each class of points has been tuned to favour the visualisation.}
		\label{fig:3dplots}
	\end{figure}
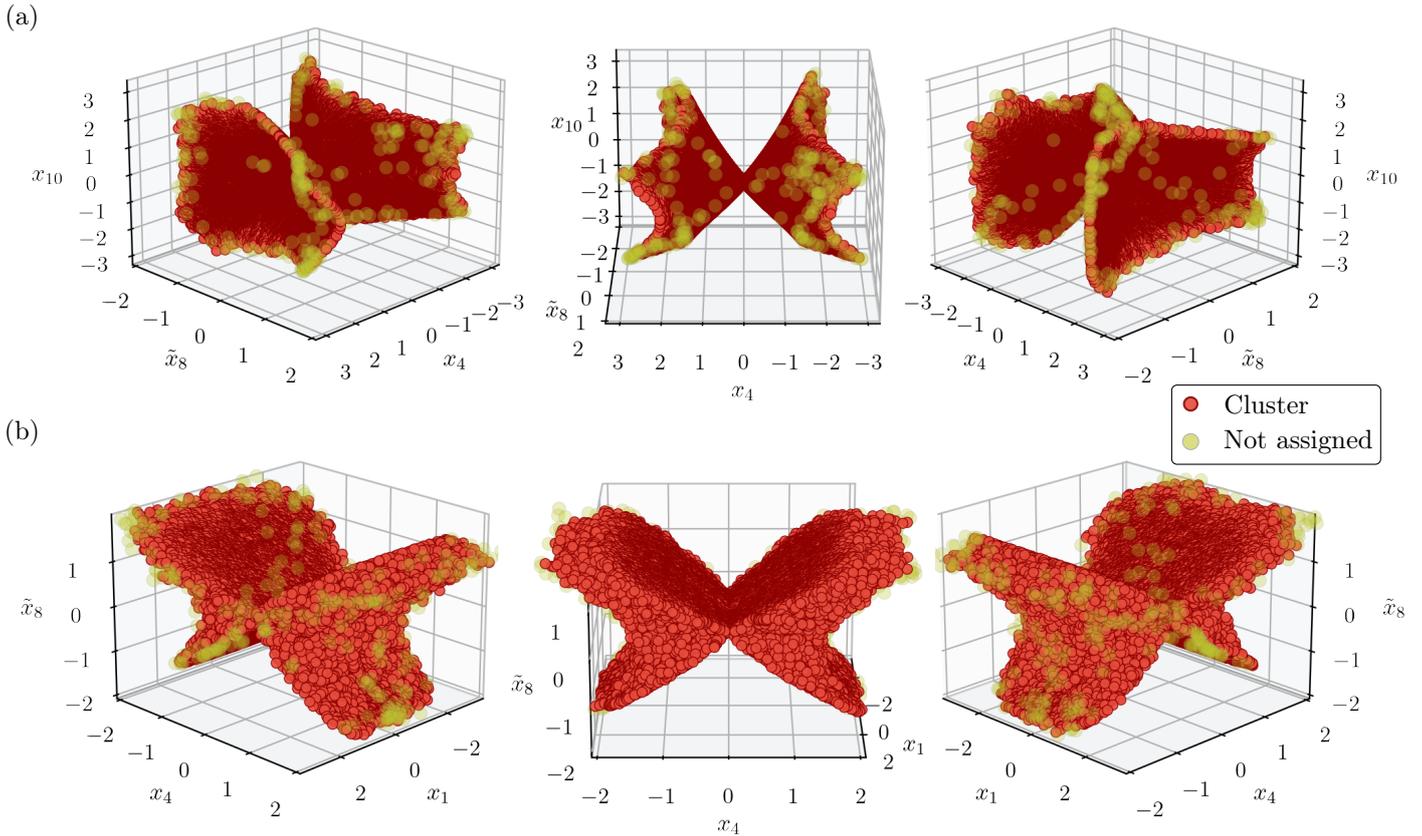

	After determining the local dimension of the space of solutions, we need to ascertain its topology. 
One possible scenario is that our data consists of several three-dimensional manifolds, intersecting at least at the origin $\vec{X}=0$, and the points previously identified with dimension $4$ may lie at the intersections of these manifolds. 
Consider for example the intersection of two lines: at the intersection point, the local dimension estimated by the previous PCA algorithm would be 2. 
To rule out this possibility, we apply a clustering algorithm on the points with tangent spaces of dimension 3 only, as identified by the PCA with $k=20$. 
This, way, we remove the possible intersection points with local dimension 4 or 5. 
For the purpose of the clustering, we use the density-based algorithm HDBSCAN~\cite{10.1007/978-3-642-37456-2_14}. 
As it can be observed in fig.~\ref{triangular_plot_1_2_4_8_10}, the points obtained after the gradient descent are denser around the origin, leading the algorithm to disregard points with a norm greater than 1. 
To avoid this effect, we de-sample the densest areas by randomly selecting no more than 10 points per 0.25-sided hyper-cube, and apply the HDBSCAN algorithm on the remaining 72,216 points, with minimum cluster size set to 5 points. 
The algorithm identifies one cluster made of 5 points, one other with 71,993 points, and fails to assign any cluster to 218 points. We show in fig.~\ref{fig:3dplots} some 3d projections of the data to visualize the clustering. In these scatter plots, red points belong to the main cluster, while yellow points are those that the algorithm failed to assign to any cluster. The smallest cluster, made of 5 points, is likely an artefact of local fluctuations in the data density and is not interpreted as physically meaningful. The size of the different points has been adjusted to facilitate the visualisation. From visual inspection, it appears that the unassigned (yellow) points lie mostly on the boundary of the sampled region. We therefore interpret their unassigned status not as evidence of belonging to another manifold, but rather as a result of insufficient local density near the edges of the dataset. Thus, the algorithm does indicate that over 99\% of the data belongs to a single dominant cluster.

	\subsection{ASMC algorithm and results}  \label{sec:AIS-SMC_application}
	From the preceding analysis, we conclude that the gradient descent procedure has produced a sampling of a single, connected, three-dimensional manifold. Therefore to characterise the manifold, we need two independent constraints on the coordinates $(x_1,x_2,x_4,\tilde{x}_8,x_{10})$. If we have a look at the form of the potential, we can observe that if we use $x_{i} = e^{\tilde{x}_i}$, then, up to some potential global factor and field redefinitions, this potential is actually a polynomial on the $\vec{X}$ coordinates. Therefore, the components of $\left.\nabla V\right|_{\vec{y}=0}$ are also polynomials on the $\vec{x} = (x_1,x_2,x_4,x_8,x_{10})$ coordinates. We conclude that the constraints on $\vec{x}$ we are looking for are polynomial constraints of the form $z(\vec{x}) = 0$, and that there should be at least two of those. Of course if one takes directly the gradient of \eqref{eq:scalarpotential}, one ends up with such conditions, but none are usable directly to solve for two of the variables in terms of the others. The problem we are facing here is therefore a problem of symbolic regression: we are looking for analytic expressions that vanish once evaluated on our data points. We now proceed to present the symbolic regression problem, utilizing the points obtained through gradient descent as our training dataset. In the following section, we detail the implementation of the ASMC framework previously outlined in sec.~\ref{Sec:AIS-SMC}.

	\subsubsection{Initialization of the algorithm}
	
	We initialized the algorithm with\footnote{$n_{\rm points}$ is the number of points used in the gradient descent, see sec.~\ref{sec:graddes} ; $n_{\rm epochs}$, $n_{\rm particles}$, $\beta_{0}$, $p_{\rm shift}$, $p_{\rm multiply}$, and $p_{\rm divide}$ are, respectively, the number of steps, the number of particles, the initial inverse temperature and the probabilities of each MCMC moves used in the ASMC procedure, see sec.~\ref{sec:detailedAISSMC} ; $\lambda$ is the regularisation parameter used in the loss function, see eq.~\eqref{eq:lossreg} ; $\sigma$ is the parameter defining the Gaussian noise for the coefficient perturbation move of the ASMC procedure, see eq.~\eqref{eq:coeffperturb}.}
	\begin{equation}	\label{eq:sample10000adaptativetemp}
		\begin{array}{lllll}
			\texttt{n\_points = 10,000}, & \qquad \qquad &  \texttt{p\_{shift} = 0.5}, & \qquad \qquad & \texttt{beta0  = 1e-6}, \\
			\texttt{n\_epochs = 1000}, & &\texttt{p\_{multiply} = 0.25}, & & \texttt{lambda = 1000}, \\
			\texttt{n\_particles = 1000}, & & \texttt{p\_{divide} = 0.25}, & & \texttt{sigma = 0.1}.\\
		\end{array}
	\end{equation}
	The convenience of these parameters was determined empirically --a systematic analysis of its optimization is left for future work. For the representation of polynomials, we adopt a vectorial approach: each polynomial is represented as an array of its coefficients $\{c_m\}$, where $m \in \{1,\dots,n_{\text{mon}}\}$ and $n_{\text{mon}}$ denotes the number of available monomials, determined by the chosen maximum degree \texttt{max\_degree} and the number of variables. We also impose a constraint on the maximum number of terms within a single polynomial \texttt{max\_num\_monomials} to promote sparsity.\footnote{An alternative approach would be to encourage sparsity through the prior distribution or within the loss function formulation.} For our simulations, we chose:
	\begin{equation} \label{eq:parampol}
		\begin{aligned}
			&\texttt{max\_degree = 4},\\
			&\texttt{max\_num\_monomials = 6.}
		\end{aligned}
	\end{equation}
With these parameters and given that we are dealing with $5$ variables, $n_{\text{mon}} = 126$.
	
Regarding the temperature schedule, as can be seen in \eqref{eq:gamman}, what matters to determine $\gamma_{n}$ is not the inverse temperature, but rather the product of the inverse temperature and the loss function. Therefore, instead of implementing an arbitrary schedule for $\beta$, we employed an adaptive temperature approach, whose aim is to ensure that the number of particles to be updated at each run is close to
\begin{equation} \label{eq:eqadaptivetemperatureschedule}
	\texttt{adaptive\_temperature\_ratio} = 0.8 - 0.5 \times \left(\frac{\texttt{n}}{\texttt{n\_epochs}}\right).
\end{equation}
In other words, if the number of accepted particles after the MCMC step at a given \texttt{n} is higher than the target~\eqref{eq:eqadaptivetemperatureschedule}, we decrease the temperature and hence increase $\beta_{n}$, and \textit{vice versa} if this number is lower than \texttt{adaptive\_temperature\_ratio}. This encourages exploration during the early stages of training and gradually transitions to a more selective regime.

	\subsubsection{Analysis of a typical run}\label{sec:1runanalysis}
	After a run, the 1000 polynomials are typically distributed into less than 50 different types of polynomials. Each particles of a given type share the same monomials, but their coefficients fluctuate a bit. We analyse them using the following procedure. We select the polynomial that features the minimal loss (without the regularisation, $\lambda=0$, see eq.~\eqref{eq:lossreg}), we fine-tune the coefficients by running a quick exploitation phase of 10,000 steps during which we randomly select a coefficient and modify it with a perturbation $\epsilon \sim \mathcal{N}(0, \sigma')$, with $\sigma'=0.01$, and keep the new polynomial only if the loss function (without the regularisation) is getting smaller. There are two possible outputs after this phase. (i)~The coefficients stay finite, and the polynomial is a candidate annihilator polynomial. We then filter out all  polynomials of the same type as this candidate, as they will ultimately give the same polynomial. Or (ii), the coefficients are just getting smaller and smaller during the exploitation phase, indicating that the only way to minimize the loss with the given monomials is to have very small coefficients. The polynomial is then disregarded. We repeat the same steps on the best polynomials of the remaining set until we have explored all polynomials.

	Here is an example for a given run. The best polynomial after the Annealing loop is
	 \begin{equation}
	 	-0.035\,x_{1} + 0.673\,x_{2} + 0.305\,x_{1}x_{4} + 0.396\,x_{1}x_{10} - 0.630\,x_{2}x_{8} - 0.284\,x_{2}x_{8}x_{10}^{2},
	 \end{equation}
	 with loss $L^{(\lambda = 0)} \simeq 2.1 \times 10^{3}$. The coefficients have been rounded to the nearest thousandth. After 10,000 steps of exploitation, this polynomial becomes
	  \begin{equation}
	 	0.396\,x_{2} + 0.396\,x_{1}x_{4} + 0.280\,x_{1}x_{10} - 0.396\,x_{2}x_{8} - 0.198\,x_{2}x_{8}x_{10}^{2},
	 \end{equation}
	 with loss $L^{(\lambda = 0)} \simeq 8 \times 10^{-5}$, or equivalently
	 \begin{equation} \label{eq:expol1}
	 	2\,x_{2} + 2.000\,x_{1}x_{4} + 1.415\,x_{1}x_{10} - 2.000\,x_{2}x_{8} - 1.000\,x_{2}x_{8}x_{10}^{2},
	 \end{equation}
	 where we normalised the coefficients by setting the one of $x_{2}$ to $2$. We removed the $x_1$ monomial, because its coefficient was smaller that $10^{-3}$. We are in case (i), the polynomial is a candidate annihilator polynomial. After filtering out the 971 polynomials that are of the same type, the next best polynomial is
	 \begin{equation}
	 	1.039\,x_{2} + 0.383\,x_{1}x_{4} + 0.466\,x_{1}x_{10} - 0.955\,x_{2}x_{8} - 0.233\,x_{2}x_{10}^{2} - 0.006\,x_{1}x_{2}x_{8}x_{10},
	 \end{equation}
	 with a loss $L^{(\lambda = 0)} \simeq 2.4 \times 10^{3}$. After the exploitation phase, it becomes
	 \begin{equation}
	 	0.006\,x_{2} + 0.004\,x_{1}x_{4} + 0.002\,x_{1}x_{10} - 0.006\,x_{2}x_{8} - 0.001\,x_{2}x_{10}^{2},
	 \end{equation}
	 with a loss $L^{(\lambda = 0)} \simeq 4 \times 10^{-2}$. The loss is low only because the coefficients are themselves very low. This is an example of case (ii), we disregard this polynomial. Reproducing similar steps for the remaining polynomials, we get some uninteresting polynomials and two new candidate polynomials (with normalised coefficients):
	 \begin{subequations}
	    \begin{align}
			1.416\,x_{2} + 1.416\,x_{1}x_{4} - 1.417\,x_{2}x_{8} + x_{1}x_{8}x_{10} - 1.001\,x_{2}x_{4}x_{8}x_{10},\quad &L^{(\lambda = 0)} \simeq 1 \times 10^{-3}, \label{eq:expol2}\\
			-1.414\,x_{1} + 1.414\,x_{1}x_{8} - 1.414\,x_{2}x_{4}x_{8} + x_{2}x_{8}x_{10},\quad &L^{(\lambda = 0)} \simeq 2 \times 10^{-4}. \label{eq:expol3}
		\end{align}
	\end{subequations}

	Given that the coefficients of the gradient $\nabla V$ are only integers and square roots of integers, the coefficients of the polynomials annihilating $\nabla V$ must be combinations of rational numbers or square roots thereof. We then deduce from eq.~\eqref{eq:expol1}, \eqref{eq:expol2} and~\eqref{eq:expol3} the candidate annihilator polynomials for the above example:
	\begin{equation}
		\begin{gathered}
			2\,x_{2} + 2\,x_{1}x_{4} + \sqrt{2}\,x_{1}x_{10} - 2\,x_{2}x_{8} - x_{2}x_{8}x_{10}^{2},\\[5pt]
	      \sqrt{2}\,x_{2} + \sqrt{2}\,x_{1}x_{4} - \sqrt{2}\,x_{2}x_{8} + x_{1}x_{8}x_{10} - x_{2}x_{4}x_{8}x_{10}, \\[5pt]
	      -\sqrt{2}\,x_{1} + \sqrt{2}\,x_{1}x_{8} - \sqrt{2}\,x_{2}x_{4}x_{8} + x_{2}x_{8}x_{10}.
		\end{gathered}
	\end{equation}

	\subsubsection{Statistics}
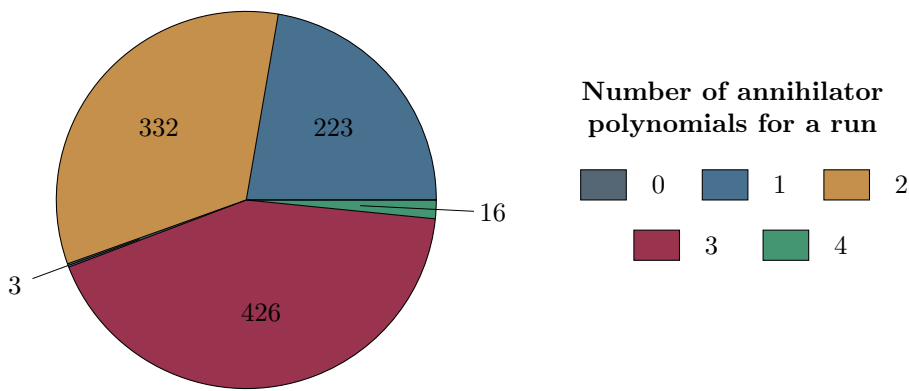
\begin{figure}[b!]
		\centering
		\begin{tikzpicture}[scale=2]

\def\radius{1.25}

\fill[myblue!80] (0:0) -- (0:\radius) arc [start angle=0,end angle={360*0.223},radius=\radius] -- (0:0);
\draw ({360*0.223/2:0.6*\radius})node [black] {\small 223};

\fill[myorange!80] (0:0) -- ({360*0.223}:\radius) arc [start angle={360*0.223},end angle={360*(0.223+0.332)}, radius = \radius] -- (0:0);
\draw ({360*(0.223+0.332/2)}:0.6*\radius) node [black] {\small 332};

\fill[myblack!80] (0:0) -- ({360*(0.223+0.332)}:\radius) arc [start angle={360*(0.223+0.332)},end angle={360*(0.223+0.332+0.003)},radius = \radius] -- (0:0);
\draw [very thin] ({360*(0.223+0.332+0.003/2)}:{\radius*0.6}) -- ({360*(0.223+0.332+0.003/2)}:\radius*1.2);
\draw ({360*(0.223+0.332+0.003/2)}:\radius*1.3) node {\small 3};  

\fill[myred!80] (0:0) -- ({360*(0.223+0.332+0.003)}:\radius) arc [start angle={360*(0.223+0.332+0.003)},end angle={360*(0.223+0.332+0.003+0.426)},radius = \radius] -- (0:0);
\draw ({360*(0.223+0.332+0.003+0.426/2)}:0.6*\radius) node [black] {\small 426};

\fill[mygreen!80] (0:0) -- ({360*(0.223+0.332+0.003+0.426)}:\radius) arc [start angle={360*(0.223+0.332+0.003+0.426)},end angle={360*(0.223+0.332+0.003+0.426+0.016)},radius = \radius] -- (0:0);
\draw [very thin] ({360*(0.223+0.332+0.003+0.426+0.016/2)}:\radius*0.6) -- ({360*(0.223+0.332+0.003+0.426+0.016/2)}:\radius*1.2);
\draw ({360*(0.223+0.332+0.003+0.426+0.016/2)}:\radius*1.3) node {\small 16};

\draw (0:0) -- (0:\radius) arc [start angle=0,end angle={360*0.223}, radius = \radius] -- (0:0);
\draw ({360*0.223}:\radius) arc [start angle={360*0.223},end angle={360*(0.223+0.332)}, radius = \radius] -- (0:0);
\draw ({360*(0.223+0.332)}:\radius) arc [start angle={360*(0.223+0.332)},end angle={360*(0.223+0.332+0.003)}, radius = \radius] -- (0:0);
\draw ({360*(0.223+0.332+0.003)}:\radius) arc [start angle={360*(0.223+0.332+0.003)},end angle={360*(0.223+0.332+0.003+0.426)},radius = \radius] -- (0:0);
\draw ({360*(0.223+0.332+0.003+0.426)}:\radius) arc [start angle={360*(0.223+0.332+0.003+0.426)},end angle={360*(0.223+0.332+0.003+0.426+0.016)}, radius = \radius] -- (0:0);

\draw (3.2,0.6) node [align=center, font= \bfseries\small] {Number of annihilator\\polynomials for a run};
\draw[black, fill=myblack!80] (2.2,0) rectangle +(0.3,0.2);
\draw (2.6,0.1) node [right] {\small $0$};
\draw[black, fill=myblue!80] (3,0) rectangle +(0.3,0.2);
\draw (3.4,0.1) node [right] {\small $1$};
\draw[black, fill=myorange!80] (3.8,0) rectangle +(0.3,0.2);
\draw (4.2,0.1) node [right] {\small $2$};
\draw[black, fill=myred!80] (2.55,-0.4) rectangle +(0.3,0.2);
\draw (2.95,-0.3) node [right] {\small $3$};
\draw[black, fill=mygreen!80] (3.4,-0.4) rectangle +(0.3,0.2);
\draw (3.8,-0.3) node [right] {\small $4$};
 
\end{tikzpicture}
		\caption{Pie chart of the repartition of the number of distinct polynomials found in a single run.}
		\label{fig:piechart}
	\end{figure}

	We performed 1000 independent runs with the same parameters as above, each involving 1000 particles. It takes approximately 10\,min to do a single run on a regular computer using CPU, including the annealing loop and the local search described in the previous section.\footnote{While writing the paper, we adapted the code to run with GPU. With a single NVIDIA Tesla P100 12 GB, it now takes 20\,s to do a single run.} The code finds the following 8 different polynomials:
	\begin{subequations} \label{eq:pols}
	 \begin{align}
	   z_{1} &= -\sqrt{2}\,x_{1} + \sqrt{2}\,x_{1}x_8 + x_{2}x_8x_{10} - \sqrt{2}\,x_{2}x_{4}x_8,\\[4pt]
	   z_{2} &= 2\,x_{2} - 2\,x_{2}x_8 + \sqrt{2}\,x_{1}x_{10} + 2\,x_{1}x_{4} - x_{2}x_8x_{10}^{2},\\[4pt]
	   z_{3} &= \sqrt{2}\,x_{2} - \sqrt{2}\,x_{2}x_8 + \sqrt{2}\,x_{1}\,x_{4} + x_{1}x_8x_{10}
	   		- x_{2}x_{4}x_8x_{10},\\[4pt]
	   z_{4} &= 2\,x_{2} - 2\,x_{2}x_8 + \sqrt{2}\,x_{1}x_8x_{10} + 2\,x_{1}x_{4}x_8
	   		- 2\,x_{2} x_{4}^{2}x_8,\\[4pt]
	   z_{5} &= -\sqrt{2}\,x_{1}^{2}x_{4} + \sqrt{2}\,x_{2}^{2}x_{4} + \sqrt{2}\,x_{1}x_{2}x_{4}^{2} - x_{1}^{2}x_{10}
	   		- x_{2}^{2}x_{10}, \\[4pt]
	   z_{6} &= -2\,x_{1} - 2\,x_{2}x_{4}- 2\,x_{1}x_{4}^{2} + 2\,x_{1}x_8 + \sqrt{2}\,x_{2}x_{10}
	   		+ x_{1}x_8x_{10}^{2},\\[4pt]
	   z_{7} &= -2 + 4\,x_{8} - 2\,x_8^2 + 2\,x_{4}^2 x_8 - x_8^2 x_{10}^2,\\[4pt]
	   z_{8} &= -2\,x_{2}x_{4} - 2\,x_{1}x_{4}^{2} + 2\,x_{2}x_{4}x_8 + \sqrt{2}\,x_{2}x_{10} 
	   		- \sqrt{2}\,x_{2}x_8x_{10} +  x_{1}x_8x_{10}^{2},
	 \end{align}
	\end{subequations}
	where $x_{8} = e^{\tilde{x}_{8}}$. They are found with different frequencies, with some polynomials occurring more often than others, as reported in tab.~\ref{table:results}. The algorithm typically identifies an average of $2.2$ distinct polynomials per run, with a maximum number of $4$, demonstrating its capacity to uncover multiple solutions simultaneously. The algorithm failed at finding a solution in only 3 runs. This is summed up in fig.~\ref{fig:piechart}. Note that in $77\%$ of the time the algorithm finds more than a single solution, demonstrating the robustness of the method.

	\begin{table}
		\centering
		\renewcommand{\arraystretch}{1.3}
		\begin{tabular}{c|ccccccccc}
			& $z_{1}$ & $z_{2}$ & $z_{3}$ & $z_{4}$ & $z_{5}$ & $z_{6}$ & $z_{7}$  & $z_{8}$&  $\varnothing$  \\\hline\hline
			Frequency & 92.6\% & 75.0\% & 51.9\% & 1.7\% & 0.2\% & 0.7\% & 0.7\% & 0.3\% & 0.1\% \\
			Maximum \# of representatives & 1000 & 1000 & 991 & 869 & 879 & 710 & 207 & 18 & --
		\end{tabular}
		\caption{Statistics of the ability of the ASMC algorithm to produce the polynomials~\eqref{eq:pols} on 1000 runs with parameters~\eqref{eq:sample10000adaptativetemp} and~\eqref{eq:parampol}. The frequencies represent the percentage of runs featuring a given polynomial in its outputs, and the maximum number of representatives gives the maximum number of particles representing the polynomial in a given run. The column $\varnothing$ counts the runs that failed to produce any annihilator polynomial.}
		\label{table:results}
	\end{table}

\begin{figure}
		\centering
		\subfigure{ \label{fig:count_pol_all_beta}
			\begin{tikzpicture}
		  		\draw (0,0) node (fig1) {\includegraphics[scale=0.7]{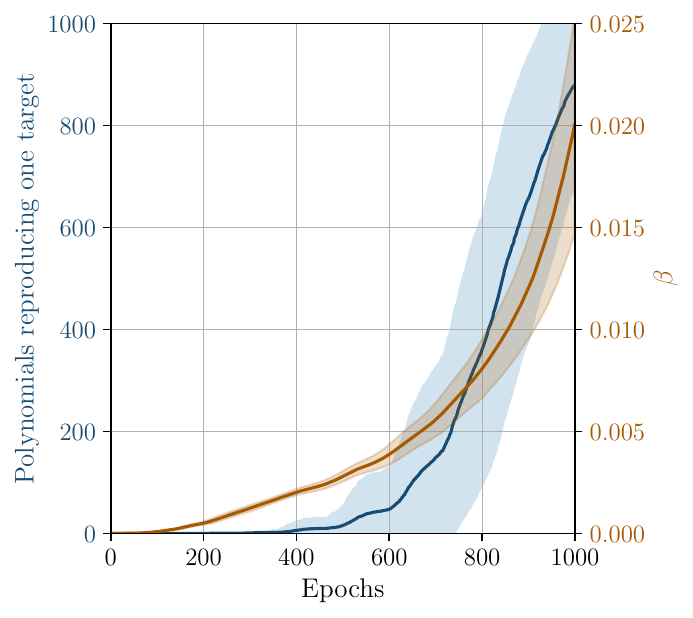}};
		  		\draw ($(fig1.north west) + (0,0)$) node {\small (a)};
		  		\draw (8.25,0) node (fig2) {\includegraphics[scale=0.7]{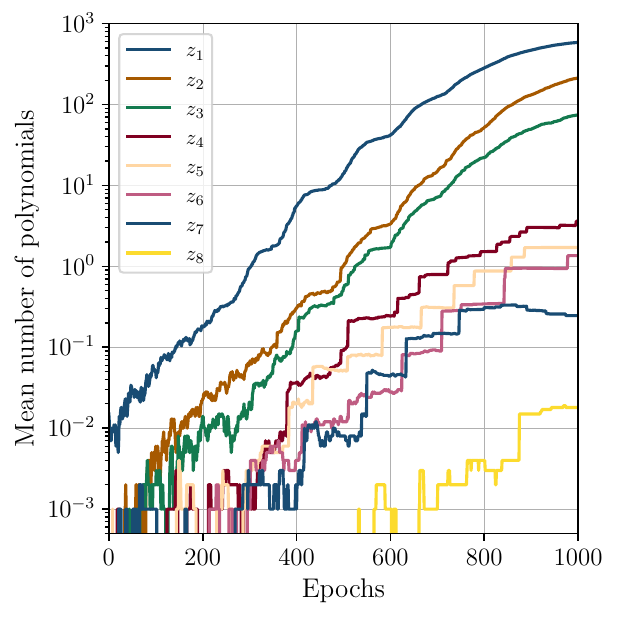}};
		  		\draw ($(fig2.north west) + (0,0)$) node {\small (b)};
		  	\end{tikzpicture}
		}
		\subfigure{\label{fig:count_pol_indiv}}
		\vspace{-1cm}
		\caption{(a) Evolution of the mean number of particles reproducing a target polynomial, along with its $1\sigma$ deviation, for the parameters described in~\eqref{eq:sample10000adaptativetemp} (left scale) and evolution of the inverse temperature $\beta$ and its $1\sigma$ deviation (right scale). (b) Dynamics of the mean number of particles reproducing each polynomial in eq.~\eqref{eq:pols} (the scale of the $y$-axis is logarithmic).}
		\label{fig:evolutionoftargetreproduction}
	\end{figure}

	We have tracked the appearance of the polynomials~\eqref{eq:pols} during each of the 1000 runs. To do so, we have counted at each epoch the number of particles including the same monomials as the annihilator polynomials.\footnote{Thus, a particle featuring the same monomials as one of the polynomials~\eqref{eq:pols} will be counted, even if the coefficients do not match, and if there are additional monomials.} The total number of particles reproducing at least one of the annihilator polynomials listed in eq.~\eqref{eq:pols} is plotted in fig.~\ref{fig:count_pol_all_beta}, together with the evolution of the inverse temperature $\beta$, both averaged on the 1000 runs. The proportion of annihilator polynomials among the particles start to be significant after approximately 700 epochs. It then increases quickly and, towards the end of each run, an average of $88\%$ of the particles do reproduce one of the annihilator polynomials, with a standard deviation of $20\%$. The evolution of this proportion is explained by the structure of the method: the code favours the particles with highest weights, and thus those that cause a significant improvement to the loss function, at each resampling. Once an annihilator polynomials is reached, it colonises larger and larger proportions of the particles at each resampling.

	The evolution of the inverse temperature is dictated by eq.~\eqref{eq:eqadaptivetemperatureschedule}. Note that $\beta$ is equal to $0.02$ on average at the end of the runs. The temperature is thus still quite high, which favours diversity. This is a key ingredient to get more than one candidate polynomials per run. The end value of $\beta$ is intimately linked to the choice of target adaptative temperature ratio in eq.~\eqref{eq:eqadaptivetemperatureschedule}: the lower this acceptance ratio, the higher the inverse temperature $\beta$. On the one hand, if the ratio is too low, very few particles get mutated and it is difficult to explore the space of polynomials and to have diverse outputs. On the other hand, if the ratio is too high, there are too many mutations: the output features a large number of polynomials, but lots of them do not converge to an annihilator polynomials because the algorithm is not selective enough. It is a matter of balance between exploration and exploitation.

	The dynamics of appearance of each annihilator polynomial is plotted in fig.~\ref{fig:count_pol_indiv}, averaged on the 1000 runs. The dynamics depend strongly on the polynomial and we observe three different classes. In the first case, for $z_{1}, z_{2}$ and $z_{3}$, the first occurrences appear typically after only few dozen of epochs; and in the second one, constituted of $z_{4}, z_{5}, z_{6}$ and $z_{7}$, after few hundreds of epochs. The last class has $z_{8}$ as its only representative. This polynomial is very difficult to produce, and when it appears it is only towards the end of the runs. The stair-step patterns, with sudden jumps in the population of the polynomials alternating with phases of stagnation, are due to the use of resampling in the algorithm. When a particle get close enough to an annihilating polynomials, its loss gets lowered significantly faster than the one of the other particles. This induces an increase of its weight and it will populate a large part of the sample at the next resampling. The coefficients get improved during the stagnation phases, inducing an increase of the polynomials weights, and thus even larger colonizations during the following resamplings. This also induces a competition between the annihilator polynomials, the ones that get bettered the more easily (typically those with fewest terms) are favoured. This mechanism explains the decrease in the averaged population of the polynomial $z_{8}$ observed after 800 epochs in fig.~\ref{fig:count_pol_indiv}.

	\begin{figure}
		\centering
		\includegraphics[scale=0.75]{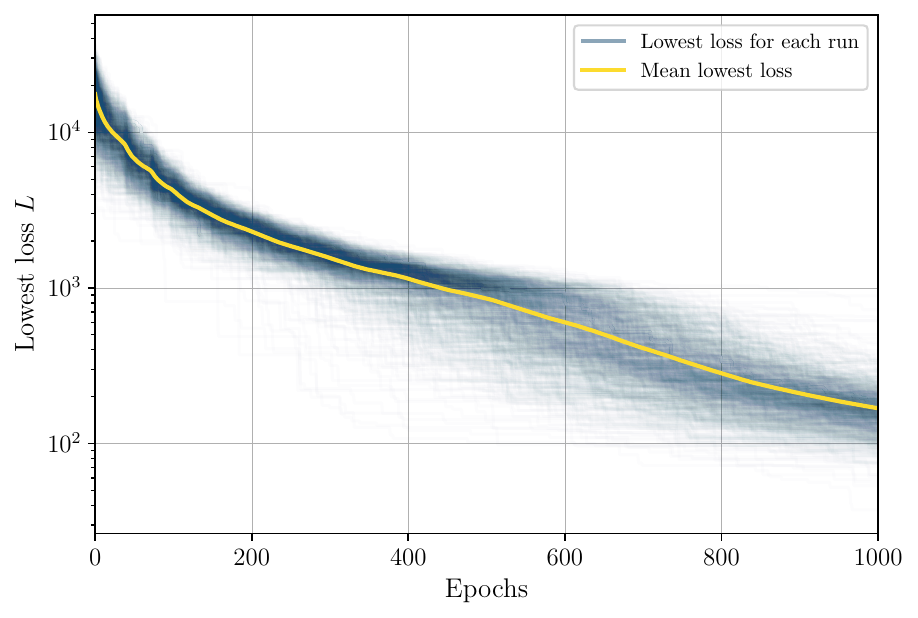}
		\caption{Evolution of the lowest loss during each of the 1000 runs (in blue) and their average (in yellow). The losses are computed with the regularisation factor $\lambda$, see eq.~\eqref{eq:lossreg}.}
		\label{fig:lossallASMC}
	\end{figure}

	The best loss for each of the 1000 runs is shown in fig.~\ref{fig:lossallASMC} in blue, together with their mean value in yellow, with a logarithmic $y$-axis. The losses here do include the regularisation factor $\lambda$, see eq.~\eqref{eq:lossreg}, and are given without the exploitation phase discussed in sec.~\ref{sec:AIS-SMC_application}. For a given run, we are plotting the best loss at each epoch, the curves thus do not follow single particles. On average the loss get bettered by two orders of magnitude during a run, illustrating the convergence of the algorithm. The best loss is however quite high at the last epoch. This is linked to the very low value of $\beta$: the exploration is favoured with respect to exploitation during all the process. Although the loss is high, 1000 epochs are sufficient to select candidate annihilator polynomials thanks to the exploitation phase performed on the best particles, as illustrated in sec.~\ref{sec:AIS-SMC_application}.

\section{Supergravity Solutions} \label{sec:sugrasol}
In the previous section, we introduced a numerical method that enabled symbolic regression, yielding a set of polynomials that vanish on our dataset, as presented in eq.~\eqref{eq:pols}. We know from sec.~\ref{sec:localanalysis} that the manifold we are aiming at parametrising is three-dimensional. As the parameter space is of dimension~$5$, we only need two constraints to define the solutions, and the eight polynomials of eq.~\eqref{eq:pols} are not independent. We are however only interested in finding an analytic parametrisation of the solutions manifold, and the easiest to find it is to solve the system
	\begin{equation}\label{eq:solvepol}
		z_i = 0, \quad \forall i \in {1,\dots,8}.
	\end{equation}
	This leads to the following rules between the parameters:
	\begin{equation}\label{eq:rulex8x10}
		\begin{cases}
			\displaystyle e^{\tilde{x}_{8}} = \frac{x_{1}^{2}+x_{2}^{2}}{x_{2}^{2} + \big(x_{1}-x_{2}x_{4}\big)^{2}},\\[10pt]
			\displaystyle x_{10} = \sqrt{2}\,x_{4}\,\frac{x_{2}^{2} - x_{1}^{2}+x_{1}x_{2}x_{4}}{x_{1}^{2}+x_{2}^{2}}.
		\end{cases}
	\end{equation}
	Alternatively, the system can be recast as:
	\begin{equation}
		\begin{cases}
			\displaystyle x_{1} = \frac{x_{2}}{e^{\tilde{x}_{8}}-1}\,\Big(x_{4}\,e^{\tilde{x}_{8}} \pm \sqrt{-1+\big(2+x_{4}^{2}\big)\,e^{x_{8}}-e^{2\tilde{x}_{8}}}\Big),\\[8pt]
			\displaystyle x_{10} = \mp\,e^{-\tilde{x}_{8}}\,\sqrt{-2+2\,\big(2+x_{4}^{2}\big)\,e^{\tilde{x}_{8}}-2\,e^{2\tilde{x}_{8}}},
		\end{cases}
	\end{equation}
	or
	\begin{equation}
		\begin{cases}
			\displaystyle x_{1} = \frac{x_{2}}{\sqrt{2}}\,\frac{e^{\tilde{x}_{8}/2}}{e^{\tilde{x}_{8}}-1}\,\Big(-x_{10}\,e^{\tilde{x}_{8}/2} \pm \sqrt{2-4\,e^{\tilde{x}_{8}}+e^{2\tilde{x}_{8}}\big(2+x_{10}^{2}\big)}\Big),\\[8pt]
			\displaystyle x_{4} = \pm \frac{e^{-x_{8}/2}}{\sqrt{2}}\,\sqrt{2-4\,e^{\tilde{x}_{8}}+e^{2\tilde{x}_{8}}\big(2+x_{10}^{2}\big)}.
		\end{cases}
	\end{equation}
	As anticipated, this defines a three-parameter manifold. We tested analytically that $\nabla V = 0$ for any of these rules, with additionnaly $\vec{y}=0$ (see eq.~\eqref{eq:defvecy}). The three-parameter manifold then defines a three-dimensional space of flat directions of the half-maximal supergravity scalar potential.\footnote{One might argue that only two of the eight polynomials are sufficient to fully characterise the solution. That is, choosing any pair $(i,j) \in {1, \dots, 8}$ may suffice to extract a complete description. In practice, this is not entirely accurate. While such a pair can yield partial constraints -- for example, recovering eq.~\eqref{eq:rulex8x10} -- it may also produce alternative (and potentially less general) parameterisations. Upon inspection, all such partial rules are found to be consistent with, and included in, the most general expressions given in eq.~\eqref{eq:rulex8x10}.}

	The supergravity solution can be shown to preserve a ${\rm U}(1)\times{\rm U}(1)$ gauged symmetry and breaks all supersymmetry, except at the origin where we recover the ${\rm SO}(4)$ isometries and the ${\cal N}=(0,4)$ supercharges. Using the parametrisation~\eqref{eq:rulex8x10}, the $(x_{1},x_{2},x_{4})$ moduli space is most nicely parametrised using the change of coordinates
	\begin{equation}
		x_{1} = r\cos(\theta)\cos(\Phi), \quad x_{2} = r\cos(\theta)\sin(\Phi) \quad {\rm and} \quad x_{4} = r\sin(\theta),
	\end{equation}
	for which the Zamolodchikov metric reads
	\begin{equation}
		\d^{2}s_{\rm Zam.} = -\,\d r^{2} - r^{2}\,\bigg(\d \theta^{2} - r\cos(\theta)\,\d \theta\d\Phi + \sin(\theta)\,\d r\d\Phi + \frac{1}{2}\,\big(3+r^{2}-\cos(2\theta)\big)\,\d\Phi^{2} \bigg).
	\end{equation}

	The spectrum of scalar fields around the flat directions is the following:
	\begin{equation}
		\begin{aligned}
			\big(m_{(0)}\ell_{\rm AdS}\big)^{2}: & \qquad 0 \ [5], \quad 8 \ [1], \quad r^{2}\big(4+r^{2}\big)\ [8],\\
			& \quad 2\,r\left(3\,r + r^{3} - r\cos(2\theta) \pm \big(2+r^{2}\big) \sqrt{2+r^{2}-2\cos(2\theta)} \right)\ [2+2].
		\end{aligned}
	\end{equation}
	The masses are normalised with respect to the AdS length $\ell_{\rm AdS}^{2} = -2/V(0) = 1/2$. The numbers between brackets indicate the multiplicity of each mode. The three-dimensional spectrum is thus stable (\textit{i.e.} satisfies the Breitenlohner-Freedman bound $\big(m_{(0)}\ell_{\rm AdS}\big)^{2}\geq -1$~\cite{Breitenlohner:1982jf}) if
	\begin{equation}
		\theta = \pm\frac{1}{2}\,\arccos\left(-1 + \sqrt{3} + \frac{- 7 + 4\sqrt{3}}{2\,r^2}\right) \quad {\rm and} \quad r \geq \sqrt{-2+\frac{7}{2\sqrt{3}}}.
	\end{equation}

\section{Conclusion} \label{sec:ccl}
We have presented a novel machine learning approach to systematically identify and characterise flat directions in supergravity scalar potentials. Our methodology combines gradient descent sampling with an Annealed Sequential Monte Carlo sampler for symbolic regression on polynomials. We demonstrated its efficiency on a 5-scalar subsector of three-dimensional half-maximal supergravity, derived from the ${\rm AdS}_3 \times S^3$ compactification of six-dimensional ${\cal N}=(1,1)$ supergravity, or analogously of the ${\rm AdS}_3 \times S^3 \times T^4$ compactification of IIB supergravity.

We developed a robust pipeline that transitions from numerical exploration to analytical understanding. The gradient descent procedure successfully samples the flat directions manifold, while local PCA analysis reveals its intrinsic dimensionality. Most notably, our Annealed Sequential Monte Carlo sampler approach to symbolic regression automatically and quickly discovers polynomial constraints characterising the manifold, bypassing the computational complexity that renders direct symbolic manipulation intractable. The algorithm uncovered eight distinct polynomial relations with varying frequencies, suggesting a hierarchical structure in constraint discovery. Furthermore, upon 1000 runs the algorithm failed at finding any solution in only 3 cases, demonstrating the robustness of the method. We compared our method to the AIFeynman algorithm~\cite{Udrescu:2019mnk} in app.~\ref{app:AIFeynman}.

This approach opens several promising avenues for advancement. The scalability to higher-dimensional cases represents the most immediate challenge and opportunity. By increasing the number of particles, refining the annealing schedule, and optimizing polynomial search strategies, we anticipate extending this methodology to the full 13-scalar theory and potentially to other supergravity models. This will pave the way to an exhaustive characterisation of the flat directions of these models, with prominent applications to the AdS/CFT correspondence: flat directions of supergravity solutions having a CFT dual are in correspondence with the space of CFT deformations preserving the conformal symmetry (called the conformal manifold). 


Our analysis reveals that the 5-dimensional scalar space contains a 3-dimensional conformal manifold. The discovered solutions preserve a ${\rm U}(1) \times {\rm U}(1)$ gauged symmetry and breaks all supersymmetries. The Zamolodchikov metric on the moduli space provides a concrete geometric description of the conformal manifold. While the specific solutions found may have limited direct physical applications, the demonstrated feasibility of our approach and its potential for systematic classification of flat directions across the supergravity landscape make it a valuable addition to the theoretical physicist's toolkit. As the solutions live in a 3d consistent truncation of both the ${\rm AdS}_3 \times S^3$ solution of ${\cal N}=(1,1)$ six-dimensional supergravity and of the ${\rm AdS}_3 \times S^3 \times T^4$ vacuum of IIB supergravity, we can uplift the solution to 6d and 10d using the tools of exceptional field theory (ExFT)~\cite{Hohm:2014fxa,Hohm:2014qga,Hohm:2017wtr}. The higher-dimensional solutions would be of the form ${\rm AdS}_3 \times M^3$ or ${\rm AdS}_3 \times M^7$, with $M^{3}$, and $M^{7}$ being deformations of the round $S^{3}$ and $S^{3}\times {\rm T}^{4}$, respectively. These deformed spaces have as moduli the 3 parameters $\{x_{1},x_{2},x_{4}\}$ that define the solutions. ExFT gives further access to the full Kaluza-Klein spectrum for those compactifications~\cite{Malek:2019eaz,Malek:2020yue,Eloy:2020uix}, allowing a test of the perturbative stability of the solutions, and giving valuable information on the spectrum of states of the dual CFT. These applications relies entirely on the fact that we have an analytic parametrisation of the solutions thanks to the symbolic regression we performed. We live these for future works.

The marriage of machine learning techniques with the study of geometry, symmetry, and dynamics in supergravity theories represents a step toward more systematic approaches to understanding the rich geometric structures underlying these theories. As computational power increases and algorithms improve, this marriage might become an important player for exploring the landscape of supergravity theories.

\section*{Acknowledgements}
We would like to thank Magdalena Larfors, whose talk on Machine Learning at the kick-off meeting of the COST action ``Fundamental challenges in theoretical physics'' in Padova in May 2024 triggered this project. We are thankful to Nelly Pustelnik, Jason Reneuve, Henning Samtleben, Dave Tennyson and Jan-Willem van de Meent for inspiring discussions. We thank especially Jan-Willem van de Meent for guiding us to the relevant numerical methods. GL was supported by endowment funds from the Mitchell Family Foundation during the initial stages of this project. The work of BD is supported by the Alexander von Humboldt foundation.

\appendix
\section{Some Details on the Supergravity Setup} \label{app:sugra}
We give here some details on the supergravity theory described schematically in sec~\ref{sec:sugra}, following ref.~\cite{Nicolai:2001ac,deWit:2003ja}.\footnote{See also ref.~\cite{Eloy:2021qol} for a review.} The Lagrangian is given in eq.~\eqref{eq: lagrangian_rephrased}. The gauging structure is encoded in an embedding tensor~\cite{Nicolai:2000sc,deWit:2002vt} that takes the general form
\begin{equation}	\label{eq: embtensor_rephrased}
	\Theta_{\bK\bL\vert\bM\bN}=\theta_{\bK\bL\bM\bN}+\frac12\Big(\eta_{\bM[\bK}\theta_{\bL]\bN}-\eta_{\bN[\bK}\theta_{\bL]\bM}\Big)+\theta\,\eta_{\bM[\bK}\eta_{\bL]\bN},
\end{equation}
where $\theta_{\bK\bL\bM\bN}=\theta_{[\bK\bL\bM\bN]}$ is fully antisymmetric, $\theta_{\bK\bL}=\theta_{(\bK\bL)}$ is symmetric and traceless, and $\theta$ is a scalar. The metric $\eta_{\bK\bL}$ is the SO(8,4)-invariant bilinear form used for index contractions. The gauge covariant derivatives are constructed using the embedding tensor according to
\begin{equation}
	D_\mu =\partial_\mu + A_\mu{}^{\bM\bN}\,\Theta_{\bM\bN\vert\bP\bQ}\, T^{\bP\bQ},
\end{equation}
where $A_\mu{}^{\bM\bN}$ are the gauge fields and
\begin{equation} \label{eq:so84gen_rephrased}
	\big(T^{\bar M\bar N}\big){}_{\bar P}{}^{\bar Q} = 2\,\delta_{\bar P}{}^{[\bar M}\,\eta^{\bar N]\bar Q}
\end{equation}
are the generators of the $\mathfrak{so}(8,4)$ algebra. The covariant derivative acting on the scalar matrix is thus
\begin{equation}
	D_\mu M_{\bM\bN}=\partial_\mu M_{\bM\bN}+4\,A_\mu{}^{\bP\bQ}\,\Theta_{\bP\bQ\vert(\bM}{}^{\bK}\, M_{\bN)\bK},
\end{equation}
ensuring gauge invariance of the scalar kinetic terms. The embedding tensor also defines the scalar potential as follows~\cite{Samtleben:2019zrh,Schon:2006kz}:
{\setlength\arraycolsep{1.2pt}
	\begin{equation}	\label{eq: scalarpot_rephrased}
		\begin{aligned}
			V	&=	\frac1{12}\,\theta_{\bK\bL\bM\bN}\theta_{\bP\bQ\bR\bS}\Big(M^{\bK\bP}M^{\bL\bQ}M^{\bM\bR}M^{\bN\bS}-6\,M^{\bK\bP}M^{\bL\bQ}\eta^{\bM\bR}\eta^{\bN\bS}\\
			&\qquad\qquad\qquad\qquad\quad+8\,M^{\bK\bP}\eta^{\bL\bQ}\eta^{\bM\bR}\eta^{\bN\bS}-3\,\eta^{\bK\bP}\eta^{\bL\bQ}\eta^{\bM\bR}\eta^{\bN\bS}\Big)\\
			&\quad +\frac1{8}\,\theta_{\bK\bL}\theta_{\bP\bQ}\Big(2\,M^{\bK\bP}M^{\bL\bQ}-2\,\eta^{\bK\bP}\eta^{\bL\bQ}-M^{\bK\bL}M^{\bP\bQ}\Big)+4\,\theta\theta_{\bK\bL}M^{\bK\bL}-32\,\theta^2.
		\end{aligned}
	\end{equation}
}
Finally, the dynamics of the vector fields is governed by the Chern-Simons contribution
\begin{equation}
	{\cal L}_{\rm CS}= -\varepsilon^{\,\mu\nu\rho}\,\Theta_{\bar M\bar N|\bar P\bar Q}\,A_{\mu}{}^{\bar M\bar N}\left(\partial_{\nu}\,A_{\rho}{}^{\bar P\bar Q}  + \frac{1}{3}\, \Theta_{\bar R\bar S|\bar U\bar V}\,f^{\bar P\bar Q,\bar R\bar S}{}_{\bar X\bar Y}\, A_{\nu}{}^{\bar U\bar V} A_{\rho}{}^{\bar X\bar Y} \right),
\end{equation}
where $f^{\bar M\bar N,\bar P\bar Q}{}_{\bar K\bar L} = 4\,\delta_{[\bar K}{}^{[\bar M}\eta^{\bar N][\bar P}\delta_{\bar L]}{}^{\bar Q]}$ are the structure constants of $\mathfrak{so}(8,4)$, and $\varepsilon^{\mu\nu\rho}$ is the three-dimensional Levi-Civita symbol.

The parametrisation of the gauging to get the truncation of six-dimensional half-maximal supergravity on $S^{3}$ is best described through the decomposition of ${\rm SO}(8,4)$ as~\cite{Eloy:2021fhc} 
\begin{equation}	\label{eq: gl3gradingbar}
	\begin{aligned}
		\text{SO}(8,4)	&\longrightarrow	\enspace \text{GL}(3,\mathbb{R})\times\text{SO}(1,1)\times\text{SO}(4)_{\rm global}\,,	\\
		X^{\bM}		&\longrightarrow	\quad \{X^{\bar m},\; X_{\bar m},\; X^\bzero,\; X_\bzero,\; X^\balpha\},
	\end{aligned}
\end{equation}
where $\bar m\in\llbracket1,3\rrbracket$ and $\balpha\in\llbracket9,12\rrbracket$ label the ${\rm SL}(3, \mathbb{R})$ and ${\rm SO}(4)_{\rm global}$ vector representations. In this basis, the ${\rm SO}(8,4)$-invariant tensor has the expression
\begin{equation}
	\eta_{\bM\bN}=
	\begin{pmatrix}
		0 & \delta_{\bar m}{}^{\bar n} & 0 & 0 &0\\
		\delta^{\bar m}{}_{\bar n} & 0 & 0 & 0 &0\\
		0 & 0 & 0 & 1 & 0\\
		0 & 0 & 1 & 0 & 0\\
		0 & 0 & 0 & 0 & -\delta_{\balpha\bbeta}
	\end{pmatrix},
\end{equation}
and the embedding tensor has the following non-vanishing components:
\begin{equation} \label{eq:embedtens}
	\theta_{\bM\bN\bP\,\bzero}=-\frac{1}{\sqrt{2}}\,X_{\bM\bN\bP}\,,		\qquad	\theta_{\bzero\bzero}=-4\sqrt{2},	
\end{equation}
with 
\begin{equation}
	X_{\bar m\bar n \bar p} = \varepsilon_{\bar m \bar n \bar p},		\qquad
	X_{\bar m}{}^{\bar n \bar p} = \varepsilon_{\bar m \bar n \bar p},	\qquad
	X^{\bar m}{}_{\bar n}{}^{\bar p} = \varepsilon_{\bar m \bar n \bar p},	\qquad
	X^{\bar m\bar n}{}_{\bar p} = \varepsilon_{\bar m \bar n \bar p}.
\end{equation}
The gauge group is then
\begin{equation}
	(T^1)^4\times[\text{SO}(4)\ltimes(T^3\times T^3)],
\end{equation}
where $T^{n}$ is a translational group transforming in the representation of dimension $n$ of ${\rm SO}(4)$. A possible parametrisation of the scalar matrix is then
\begin{equation}	\label{eq: scalarmatrix}
	M_{\bM\bN}
	=
	\begin{pmatrix}
		m+(\xi^2+\phi)m^{-1}(\xi^2-\phi)+2\xi^2	& 	(\xi^2+\phi)m^{-1}		&	0	&  0  &	-\sqrt2\,[1+(\xi^2+\phi)m^{-1}]\xi	\\
		m^{-1}(\xi^2-\phi)	& 	m^{-1}	&	0  	&  0  &	-\sqrt2\,m^{-1}\xi	\\
		0		&	0		&	e^{2\tilde\varphi}	&	0				&	0		\\
		0		&	0		&	0				&	e^{-2\tilde\varphi}	&	0		\\
		-\sqrt2\,\xi^T[1+m^{-1}(\xi^2-\phi)]	&	-\sqrt2\,\xi^Tm^{-1}	&	0	&	0	&	1+2\,\xi^Tm^{-1}\xi	\\
	\end{pmatrix},
\end{equation}
in terms of a symmetric ${\rm GL}(3,\mathbb{R})$ matrix $m$, a $3\times3$ antisymmetric matrix $\phi$, a $3\times4$ matrix $\xi$, its $3\times 3$ square $\xi^{2} = \xi \xi^{T}$ and a dilaton $\tilde{\varphi}$. This encodes 22 out the 32 scalars of the theory, 10 of the scalars being gauge fixed using the translations in the gauge group. With this parametrisation, the potential takes the form~\eqref{eq:scalarpotential}. We can further consistently restrict ourselves to a set of 13 scalars by requiring invariance under the diagonal ${\rm SO}(3)$ subgroup of ${\rm SO}(4)_{\rm global}$~\cite{Eloy:2021fhc}, yielding to the parametrisation described in eq.~\eqref{eq:13scalars1} and~\eqref{eq:13scalars2}.

\section{Comparison with AIFeynman} \label{app:AIFeynman}
We have used the state-of-the-art technique \texttt{AIFeynman} \cite{Udrescu:2019mnk} on our dataset, with again $x_8 = \exp(\tilde{x}_8)$, so as to identify polynomial expressions fitting the data. We have run it with the following configuration 
\begin{equation}
	\begin{aligned}
		&\texttt{BF\_try\_time} = 60, \\
		&\texttt{polyfit\_deg} = 5, \\
		&\texttt{NN\_epochs} = 1000.\\
	\end{aligned}
\end{equation}
The first parameter fixes the time limit in seconds for each brute force call, \texttt{polyfit\_deg} gives the maximum degree of the polynomial tried by the polynomial fit routine and \texttt{NN\_epochs} is the number of training epoch for the internal neural network. The function used for the brute force tests are 
\begin{equation}
\texttt{+*-/><} \sim \texttt{\textbackslash R1}.
\end{equation}
The binary operations are addition, multiplication, substraction and division. The unary ones are inverse, increment, decrement, negation and square root. Finally there is a nonary one, the unity. For more details we refer to \cite{Udrescu:2019mnk}.

In this algorithm, one tries to fit one of the variables in term of the others. Here, we used it to fit $x_{10}$ in terms of the others. The AIFeynman algorithm finds a solution: 
\begin{equation}
	\texttt{x\_10 = 1.414213551821*(x\_4-((x\_1/x\_2)-((x\_1/x\_2)/x\_8))},
\end{equation}
which after identifying the numerical factor with $\sqrt 2$ and inverting the relation gives 
\begin{equation}
	- \sqrt 2\, x_1 + \sqrt 2\, x_1 x_8 + x_2 x_8  x_{10} -\sqrt 2\, x_2 x_4 x_8 = 0,
\end{equation}
which corresponds to $z_1$ of eq.~\eqref{eq:pols}. While the AIFeynman code is able to recover one of the constraints, it exhibits several drawbacks compared to our method. First, it is significantly slower: it required 5685 seconds (1 hour 34 minutes and 45 seconds) compared to approximately 600 seconds for a single run of our ASMC algorithm.\footnote{This time only accounts for the annealing loop, without the local search algorithm described in sec.~\ref{sec:1runanalysis}. Furthermore, this computational time reflects the overhead incurred by verifying at each iteration whether the particles reproduce one of the polynomials specified in eq.~\eqref{eq:pols}, which substantially increases the overall execution time. In the absence of this verification step, the annealing loop would require approximately 350 seconds on standard computational hardware.} Moreover, the AIFeynman framework is based on expressing one variable as a function of the others, which assumes the invertibility of the underlying relation. In general, this is not guaranteed for the class of polynomials in eq.~\eqref{eq:pols}, and obtaining a closed-form inverse can be nontrivial or even impossible for higher-degree expressions. In our case, each polynomial involves at most quadratic terms in any single variable, ensuring invertibility, but this property would not hold for more complex models. In addition, AIFeynman is limited to recovering one constraint per run, whereas our ASMC method can discover multiple constraints simultaneously. This is reflected in our results, where each run identified an average of 2.2 polynomials, with up to 4 found in the best cases. Finally, the brute-force regression strategy used by AIFeynman makes it less effective for identifying higher-degree polynomials, whose complexity increases the search difficulty. In contrast, our method treats all polynomials as equally probable, regardless of their degree, enabling it to uncover more intricate structures more efficiently.\footnote{Note that we do not claim that we have used the AIFeynman in the most optimal way.}

\bibliography{references}

\end{document}